\documentclass[11pt,a4paper]{article}
\usepackage[colorlinks=false]{hyperref}

\usepackage{mathrsfs, bm, amsmath, amsthm, amssymb, xcolor}
\usepackage{soul}
\usepackage{graphicx}
\usepackage[english]{babel}
\usepackage{booktabs, multirow}
\usepackage{pdflscape}
\usepackage{array}  
\makeatletter
\usepackage{appendix}

\renewcommand\eqref[1]{Eq.~(\ref{eq:#1})}
\newcommand\eqrefp[1]{(\ref{eq:#1})}
\AtBeginDocument{\providecommand\secref[1]{Section~\ref{sec:#1}}}
\AtBeginDocument{\providecommand\subsecref[1]{Section~\ref{subsec:#1}}}
\AtBeginDocument{\providecommand\figref[1]{Figure~\ref{fig:#1}}}
\AtBeginDocument{\providecommand\tabref[1]{Table~\ref{tab:#1}}}

\providecommand{\tabularnewline}{\\}

\theoremstyle{plain}
\newtheorem*{thm*}{\protect\theoremname}
\theoremstyle{plain}

\usepackage{accents}
\newlength{\dhatheight}

\definecolor{darkgreen}{RGB}{0 155 0}

\newcommand\ie{\textit{i.e.}}
\newcommand\eg{\textit{e.g.}}

\newcommand{\abs}[1]{\left| #1 \right|}

\makeatother

\providecommand{\theoremname}{Theorem}

\usepackage{natbib}
\usepackage[left=25mm,right=25mm,top=1.5cm,bottom=1.5cm,includeheadfoot]{geometry}
\begin{document}
	
\newcommand\cm{\mathrm{\,cm}}
\newcommand\mm{\mathrm{\,mm}}
\newcommand\m{\mathrm{m}}
\newcommand\cdf{\mathrm{CDF}}
\newcommand\pdf{\mathrm{PDF}}
\newcommand\AIC{\mathrm{AIC}}
\newcommand\card{\mathrm{card}}
\newcommand\X{\bm{X}}
\newcommand\Xind{\bm{X^{\prime}}}
\newcommand\Z{\bm{Z}}
\newcommand\x{\bm{x}}
\renewcommand\u{\bm{u}}
\newcommand\fX{f_{\bm{X}}}
\newcommand\FX{F_{\bm{X}}}
\newcommand\CX{C_{\bm{X}}}
\newcommand\cX{c_{\bm{X}}}
\newcommand\FXext{(F_{1}(x_{1}),\,\ldots,\,F_{M}(x_{M}))} 
\newcommand\FXinvext{(F_{1}^{-1}(u_{1}),\,\ldots,\,F_{M}^{-1}(u_{M}))} 
\newcommand\xext{(x_{1},\,\ldots,\,x_{M})} 
\newcommand\uext{(u_{1},\,\ldots,\,u_{M})} 
\newcommand\Y{\bm{Y}}
\newcommand\y{\bm{y}}
\newcommand\z{\bm{z}}
\newcommand\fY{f_{Y}}
\newcommand\R{\mathcal{T}^{(\mathcal{R})}}
\newcommand\Rinv{(\mathcal{T}^{(\mathcal{R})})^{-1}}
\newcommand\Rtilde{\hat{\mathcal{T}^{(\mathcal{R})}}}
\newcommand\Rtildeinv{(\hat{\mathcal{T}^{(\mathcal{R})}})^{-1}}

\newcommand\tr{\mathrm{T}}
\newcommand{\corrmat}{\mathbf{R}}
\newcommand{\idmat}{\mathbf I}
\newcommand\M{\mathcal{M}}
\newcommand\N{\mathcal{N}}
\newcommand\Vine{\mathcal{V}}
\newcommand\D{\mathcal{D}}
\newcommand\DX{\D_{\X}}
\newcommand\RR{\mathbb{R}}
\newcommand\RRM{\mathbb{R}^{M}}
\newcommand\A{\mathcal{A}}
\newcommand\notA{\overline{\mathcal{A}}}
\newcommand\T{\mathcal{T}^{(\mathcal{N})}}
\newcommand\TU{\mathcal{T}^{(\mathcal{U})}}
\newcommand\TP{\mathcal{T}^{(\Pi)}}
\newcommand\TPR{\mathcal{T}^{(\Pi,\mathcal{R})}}
\newcommand\TPRa{\mathcal{T}_{1}^{(\Pi,\,\mathcal{R})}}
\newcommand\TPRb{\mathcal{T}_{2}^{(\Pi,\,\mathcal{R})}}
\newcommand\TPNa{\mathcal{T}_{1}^{(\Pi,\,\mathcal{N})}}
\newcommand\Ran{\textrm{Ran}}
\newcommand\Tinv{(\mathcal{T}^{(\mathcal{N})})^{-1}}
\newcommand\TG{\mathcal{T}^{(G)}}
\newcommand\G{\bm{G}}
\renewcommand\P{\mathbb{P}}

\newcommand\E{\mu}
\newcommand\V{\sigma^2}
\newcommand\std{\sigma}

\newcommand\Cgauss{C^{(\mathcal{N})}}
\newcommand\Cgumb{C^{(\mathcal{GH})}}
\newcommand\FcondA{F_{\notA|\A}}
\newcommand\fcondA{f_{\notA|\A}}
\newcommand\FicondA{F_{i|\A}}
\newcommand\ficondA{f_{i|\A}}
\newcommand\CcondA{C_{\notA|\A}}
\newcommand\ccondA{c_{\notA|\A}}
\newcommand\Pa{\,\mathrm{N}}
\newcommand\kN{\,\mathrm{kN}}
\newcommand\Psialpha{\Psi_{\bm{\alpha}}}
\newcommand\Psibeta{\Psi_{\bm{\beta}}}
\newcommand\phialpha{\phi_{\alpha_{i}}^{(i)}}
\newcommand\phibeta{\phi_{\beta_{i}}^{(i)}}
\newcommand\Lfix{\mathrm{L}_{\mathrm{fix}}}
\newcommand\Lext{\mathrm{L}_{\mathrm{ext}}}
\newcommand\Ltot{\mathrm{L}_{\mathrm{tot}}}
\newcommand\CInd{C^{(\Pi)}}
\newcommand\CGauss{C^{(\N)}}
\newcommand\CVine{C^{(\Vine)}}
\newcommand\cVine{c^{(\Vine)}}
\newcommand\CVineHat{\hat{C}^{(\Vine)}}

\newcommand\muMC{\hat{\mu}_{\textrm{MC}}}
\newcommand\sigmaMC{\hat{\sigma}_{\textrm{MC}}}
\newcommand\muIndMC{\hat{\mu}_{\textrm{MC}}^{(\Pi)}}
\newcommand\sigmaIndMC{\hat{\sigma}_{\textrm{MC}}^{(\Pi)}}
\newcommand\muGaussMC{\hat{\mu}_{\textrm{MC}}^{(\N)}}
\newcommand\sigmaGaussMC{\hat{\sigma}_{\textrm{MC}}^{(\N)}}
\newcommand\muVineMC{\hat{\mu}_{\textrm{MC}}^{(\Vine)}}
\newcommand\sigmaVineMC{\hat{\sigma}_{\textrm{MC}}^{(\Vine)}}
\newcommand\muVineHatMC{\hat{\mu}_{\textrm{MC}}^{(\hat{\Vine})}}
\newcommand\sigmaVineHatMC{\hat{\sigma}_{\textrm{MC}}^{(\hat{\Vine})}}

\newcommand\muPCE{\hat{\mu}_{\textrm{PCE}}}
\newcommand\sigmaPCE{\hat{\sigma}_{\textrm{PCE}}}
\newcommand\muIndPCE{\hat{\mu}_{\textrm{PCE}}^{(\Pi)}}
\newcommand\sigmaIndPCE{\hat{\sigma}_{\textrm{PCE}}^{(\Pi)}}
\newcommand\muGaussPCE{\hat{\mu}_{\textrm{PCE}}^{(\N)}}
\newcommand\sigmaGaussPCE{\hat{\sigma}_{\textrm{PCE}}^{(\N)}}
\newcommand\muVinePCE{\hat{\mu}_{\textrm{PCE}}^{(\Vine)}}
\newcommand\sigmaVinePCE{\hat{\sigma}_{\textrm{PCE}}^{(\Vine)}}
\newcommand\muVineHatPCE{\hat{\mu}_{\textrm{PCE}}^{(\hat{\Vine})}}
\newcommand\sigmaVineHatPCE{\hat{\sigma}_{\textrm{PCE}}^{(\hat{\Vine})}}

\newcommand\PfIndMC{\hat{P}_{f;\textrm{MC}}^{(\Pi)}}
\newcommand\PfGaussMC{\hat{P}_{f;\textrm{MC}}^{(\N)}}
\newcommand\PfVineMC{\hat{P}_{f;\textrm{MC}}^{(\Vine)}}
\newcommand\PfVineHatMC{\hat{P}_{f;\textrm{MC}}^{(\hat{\Vine})}}

\newcommand\PfIndFORM{\hat{P}_{f;\textrm{FORM}}^{(\Pi)}}
\newcommand\PfGaussFORM{\hat{P}_{f;\textrm{FORM}}^{(\N)}}
\newcommand\PfVineFORM{\hat{P}_{f;\textrm{FORM}}^{(\Vine)}}
\newcommand\PfVineHatFORM{\hat{P}_{f;\textrm{FORM}}^{(\hat{\Vine})}}

\newcommand\PfIS{\hat{P}_{f;\textrm{IS}}}
\newcommand\PfIndIS{\hat{P}_{f;\textrm{IS}}^{(\Pi)}}
\newcommand\PfGaussIS{\hat{P}_{f;\textrm{IS}}^{(\N)}}
\newcommand\PfVineIS{\hat{P}_{f;\textrm{IS}}^{(\Vine)}}
\newcommand\PfVineHatIS{\hat{P}_{f;\textrm{IS}}^{(\hat{\Vine})}}

\newcommand\StdVinePCE{\tilde{\sigma}^{(\Vine)}} 
\newcommand\EVinePCE{\tilde{\mu}^{(\Vine)}}

\newcommand\uqlab{\textsc{UQLab}}

\title{A general framework for data-driven uncertainty quantification under complex input dependencies using vine copulas}
\author{E. Torre, S. Marelli, P. Embrechts, B. Sudret}
\maketitle

\abstract{
Systems subject to uncertain inputs produce uncertain responses. Uncertainty 
quantification (UQ) deals with the estimation of statistics of the system 
response, given a computational model of the system and a probabilistic model 
of its inputs. In engineering applications it is common to assume that the 
inputs are mutually independent or coupled by a Gaussian or elliptical 
dependence structure (copula).

In this paper we overcome such limitations by modelling the dependence structure of multivariate inputs as vine copulas. Vine copulas are models of multivariate dependence built from simpler pair-copulas. The vine representation is flexible enough to capture complex dependencies. This paper formalises the framework needed to build vine copula models of multivariate inputs and to combine them with virtually any UQ method. The framework allows for a fully automated, data-driven inference of the probabilistic input model on available input data. 

The procedure is exemplified on two finite element models of truss structures, both subject to inputs with non-Gaussian dependence structures. For each case, we analyse the moments of the model response (using polynomial chaos expansions), and perform a structural reliability analysis to calculate the probability of failure of the system (using the first order reliability method and importance sampling). Reference solutions are obtained by Monte Carlo simulation. The results show that, while the Gaussian assumption yields biased statistics, the vine copula representation achieves significantly more precise estimates, even when its structure needs to be fully inferred from a limited amount of observations. 

\vspace*{1em}
}

\section{Introduction}
\label{sec:Introduction}

Uncertainty Quantification (UQ) estimates statistics of the response of a system subject to stochastic inputs. The system is usually described by a deterministic computational model $\M$ (\eg, a finite element code). The input consists of $M$ possibly coupled parameters, modelled by a random vector $\X$ with joint cumulative distribution function ($\cdf$) $\FX$ and probability density ($\pdf$) $\fX$. The computational model transforms $\X$ into an uncertain output $Y=\M(\X)$, which here we take to be a univariate random variable. The extension to multivariate outputs is straightforward. 

Of interest in UQ problems are various statistics of $Y$, such as its
$\cdf$ $F_{Y}$, its moments, the probability of extreme events (\ie, of
small or large quantiles), the sensitivity of $Y$ to the different
components $X_{i}$ of $\X$, and others. Because $\M$ is typically a
complex model which is not known explicitly, analytical solutions are in
general not available. The model behavior can only be known point-wise
in correspondence with inputs $\x^{(j)}$ sampled from $\FX$, where it
produces responses $y^{(j)} = \M(\x^{(j)})$ (non-intrusive, or black-box
approach). The classical and most general strategy to solve this class
of problems is by Monte Carlo simulation (MCS).  MCS draws the
$\x^{(j)}$ as i.i.d samples from $\FX$, which requires the sample size
$n$ to be large enough to cover the input probability space sufficiently
well. When $\M$ is computationally expensive and the available
computational budget is limited to a few dozens to hundreds of runs,
alternative approximation techniques are used instead of MCS. Examples
include the first and second order reliability methods (FORM
\citep{Hasofer1974}, SORM \citep{Fiessler1979_SORM}), importance sampling
(IS, \cite{Melchers1999}) and subset simulation \citep{Au2001} in
reliability analysis for the estimation of small failure probabilities
(see also \cite{Ditlevsen1996, Lemaire09}), and polynomial chaos
expansions (PCE, \cite{Ghanem98}), Kriging \citep{Matheron67}, and other
metamodelling techniques for the estimation of the moments.

Since $\M$ is a deterministic code, all uncertainty in $Y$ is due to the
uncertainty in $\X$. Therefore, regardless of the approach (MCS or
others) chosen to estimate the statistics of $Y$ of interest, a suitable
model of $\FX$ is critical to obtain accurate estimates. Historically,
the components $X_{i}$ of $\X$ are assumed to be mutually independent,
or to have the dependence structure of a multivariate elliptical
distribution \citep{Lebrun2009c}. Among the latter, Gaussian
distributions are often employed because they are simple to model and to
fit to data, since they only require the computation of pairwise
correlation coefficients. In addition, some advanced UQ techniques take
advantage of (or require) mutually independent inputs. These include
FORM, SORM, IS, some types of subset simulation (\eg,
\cite{Papaioannou2015_89}), PCE. The most general transformation to map
the input vector $\X$ onto a vector $\Z$ with independent components,
the Rosenblatt transform \citep{Rosenblatt52}, requires the computation
of conditional $\pdf$s, which are hardly known in practical
applications. However, when $\FX$ has a Gaussian dependence structure,
this map is known and is equivalent to the well known Nataf transform
\citep{Nataf62, Lebrun2009c}. The Gaussian assumption introduces thus a
convenient representation of input dependencies. When the real
dependence structure deviates from this assumption, it may however
introduce a bias in the resulting estimates. The validity or the impact
of the Gaussian assumption, though, are typically not quantified. Novel
methodologies in UQ largely focus on providing better estimation
techniques rather than on allowing for different probabilistic input
models.

Recently, dependence modelling has seen significant advances in the
mathematical community with the widespread adoption of copula models,
and of vine copulas in particular. Copula theory allows to separately
model the dependence (by multivariate copula functions) and the marginal
behaviour (by univariate $\cdf$s) of joint distributions. This provides
a flexible way to build multivariate probability models by selecting
each ingredient individually \citep{Nelsen2006,Joe2015}. Copulas have
recently been used in various studies in engineering, such as in
earthquake \citep{Goda2010_112,Goda2015_39,Zentner2017_54} and sea waves
\citep{Michele2007_734, Masina2015_97, MontesIturrizaga2016_564}
engineering. Applications, however, are often limited to low-dimensional
(typically bivariate) problems, or to relatively simple copula families,
prominently the Gaussian or Archimedean families \citep{Nelsen2006}. In
higher dimensions, building and selecting copulas that properly
represent the coupling of the phenomena of interest may be a complex
problem. Vine copulas, first established by \citet{Joe1996} and
\citet{Bedford2002}, ease this construction by expressing multivariate
copulas as a product of simpler bivariate copulas among pairs of random
variables. As a result, vine models offer an easy interpretation and are
extremely flexible. Vine copulas have been extensively employed, for
instance, in financial applications \citep{Aas2016}. In engineering,
these models have been, so far, largely overseen. Recently,
\citet{Wang2017_1, Wang2017_22} proposed their application in the
context of reliability analysis, for the special case when only partial
information (correlation coefficients) is available. In a later study,
they used vine copulas in combination with MCS for reliability analysis
\citep{Wang2017_inpress}.

This manuscript proposes a general framework to use vine copulas to
model model input dependencies in UQ problems. The flexibility of these
models guarantees an accurate description of the input dependence
properties that shape the output statistics. Besides, since algorithms
to compute the Rosenblatt transform of vine copulas are available, these
dependence models are applicable also in combination with UQ techniques
that work in probability spaces with independent variables. Algorithms
to infer the structure and fit the parameters of vine models to data,
for instance based on maximum likelihood or Bayesian estimation, also
exist, making these models suitable for data driven applications
\citep{Aas2009, Schepsmeier2015}.

After recalling fundamental results of copula and vine copula theory (Sections \ref{sec:Copulas-and-vines}-\ref{sec:vines4UQ}), we combine three established UQ methodologies, FORM, IS and PCE, with vine copula models of the input dependencies (\secref{UQ-techniques}). In Sections \ref{sec:horizontal_truss}-\ref{sec:dome} we apply the methodology to two truss models. We show that modelling non-Gaussian input dependencies with the Gaussian copula yields wrong estimates of the failure probability and of the response moments. The problem cannot be amended by using different UQ methods, since it is inherent to the wrong representation of the input uncertainty. Reliable estimates are obtained instead by using a suitable vine representation of the input, also when the vine is purely inferred from available data. The method's advantages and current limitations are discussed in \secref{discussion}.

\section{Copulas and vine copulas}
\label{sec:Copulas-and-vines}

Multivariate inputs in UQ problems are generally modelled as random vectors. The statistical properties of an $M$-dimensional random vector $\X$ are fully described by its joint $\cdf$ 
\[
\FX(\x)=\P(X_1 \leq x_1,\, \ldots, X_M \leq x_M).
\]

The joint $\cdf$ defines both the marginal $\cdf$ of each component $X_{i}$ of $\X$, \ie, $F_{i}(x_{i})=F_{X_{i}}(x_{i})=\P(X_{i} \leq x_{i})$, $i=1,\ldots,M$,
and the dependence properties of the variables. As such, prescribed parametric families of joint $\cdf$s dictate specific parametric forms for the marginal and joint properties of the random variables. More flexible models should be compatible with inference techniques, to be applicable when only a finite number of realisations of the input $\X$ is available. They should also optimally provide the isoprobabilistic map that decouples their random variables, such to be usable in combination with UQ techniques that assume mutually independent inputs. This section introduces vine copula models and illustrates how they meet the requirements listed above. 

\subsection{Copulas and Sklar's theorem}
\label{subsec:Copulas-and-Sklars-theorem}

An $M$-copula is defined as an $M$-variate joint $\cdf$ $C:[0,\:1]^{M}\rightarrow[0,1]$ with standard uniform marginals, that is, such that
\[
C(1,\ldots,1,u_{i},1,\ldots,1)=u_{i}\quad\forall u_{i}\in[0,1],\quad\forall i=1,\ldots,M.
\]

Sklar's theorem \citep{Sklar1959} allows one to express joint $\cdf$s in terms of their marginal distributions and a copula.

\begin{thm*}[\textbf{Sklar}]
For any $M$-variate $\cdf$ $\FX$ with marginals $F_{1},\ldots,\:F_{M}$, an $M$-copula $\CX$ exists, such that for all $\x \in \RRM$
\begin{equation}
\FX(\x)=\CX\FXext.
\label{eq:Sklar-F-C-relation}
\end{equation}
Besides, $\CX$ is unique on $\Ran(F_{1})\times\ldots\times\Ran(F_{M})$, where $\Ran$ is the range operator. In particular, $\CX$ is unique on $[0,\,1]^{M}$ if all $F_{i}$ are continuous, and it is given by
\begin{equation}
\CX(\u)=\FX\FXinvext, \quad \u \in [0,1]^M.
\label{eq:Sklar-C-from-F}
\end{equation}
Conversely, for any $M$-copula $C$ and any set of $M$ univariate
$\cdf$s $F_{i}$ with domain $\D_{i}$, $i=1,\ldots,\,M$, the function
$F:\,\D_{1}\times\ldots\times\D_{M}\rightarrow[0,\,1]$ defined by
\begin{equation}
F\xext:=C\FXext
\label{eq:Sklar-F-C-relation2}
\end{equation}
is an $M$-variate $\cdf$ with marginals $F_{1},\ldots,F_{M}$. 
\end{thm*}

The representation \eqrefp{Sklar-F-C-relation} guarantees that any joint $\cdf$ can be expressed in terms of its marginals and a copula. In the following we work with joint $\cdf$s $\FX$ having continuous marginals $F_{i}$.

Copulas of known families of joint $\cdf$s can be derived from \eqrefp{Sklar-C-from-F}. Finally, one can use \eqrefp{Sklar-F-C-relation2} to build a multivariate $\cdf$ $F$ by separately specifying and combining $M$ univariate $\cdf$s $F_{i}$ and a copula $C$. The univariate $\cdf$s describe the marginal behaviour, while the copula describes the dependence properties. Sklar's theorem thus allows one to split the problem of modelling the joint behaviour of the components of $\X$ into two separate problems. One first models the marginals $F_{i}$, then transforms the original components $X_{i}$ into uniform random variables $U_{i}=F_{i}(X_{i})$, leading to the transformation
\begin{equation}
\mathcal{T}^{(\mathcal{U})}: \X \mapsto \bm{U} = \left( F_1(X_1),\,\ldots,F_M(X_M) \right)^\tr.
\label{eq:Uniform-transformation}
\end{equation}
The joint $\cdf$ of $\bm{U}=(U_{1},\ldots,\,U_{M})^\tr$ is the associated copula. 

Sklar's theorem can be re-stated in terms of probability densities. If $\X$ admits $\pdf$ $\displaystyle{\fX(\x):=\frac{\partial^M \FX(\x)}{\partial x_1\ldots \partial x_M}}$ and copula density $\displaystyle{\cX(\u):=\frac{\partial^M \CX(\u)}{\partial u_1\ldots \partial u_M}}$, then the following relation holds:
\begin{equation}
\fX(\x)=\cX\FXext\cdot\prod_{i=1}^{M}f_{i}(x_{i}).
\label{eq:f-c-rel}
\end{equation}

\subsection{Copula-based measures of dependence}
\label{subsec:Copula-measures-of-dep}

Since copulas fully describe multivariate dependencies, it is natural to introduce dependence measures based on the copula only, and not on the marginals. Several such measures, also known as measures of concordance, exist. An example is Spearman's correlation coefficient, defined for a random pair $(X_{1},\,X_{2})$ as
\[
\rho_S(X_1, X_2) := \rho_P(F_1(X_1), F_2(X_2)),
\]
where $\rho_P$ is the classical Pearson correlation coefficient. Another example is Kendall's tau
\[
\tau_{K}(X_1, X_2):=\P((X_{1}-\tilde{X}_{1})(X_{2}-\tilde{X}_{2})>0)-\P((X_{1}-\tilde{X}_{1})(X_{2}-\tilde{X}_{2})<0),
\]
where $(\tilde{X_{1}},\,\tilde{X_{2}})$ is an independent copy of $(X_{1},\,X_{2})$. If the copula of $(X_{1},\,X_{2})$ is $C$, then 
\begin{equation}
\rho_{S}(X_1, X_2)=12\iint_{[0,1]^{2}}C(u,\,v)du dv-3 = 3 - 12\iint_{[0,1]^2} u \frac{\partial C(u,v)}{\partial u},
\label{eq:Spearman-rho-from-C}
\end{equation}
and
\begin{equation}
\tau_{K}(X_1, X_2)=4\iint_{[0,1]^{2}}C(u,\,v)dC(u,v)-1= 1 - 4 \iint_{[0,1]^2} \frac{\partial C(u,v)}{\partial u} \frac{\partial C(u,v)}{\partial v} du dv,
\label{eq:Kendalls-tau-from-C}
\end{equation}
where the RHS in both equations is well defined if the copula partial derivatives exist and are not degenerate at the borders \citep{Joe2015}.

One can show that $\tau_{K}=0$ and $\rho_S=0$ if $(X_{1},\,X_{2})$ are
independent, that $\tau_{K}=1 \Leftrightarrow \rho_S=1 \Leftrightarrow
X_{1}=\alpha(X_{2})$ for some strictly increasing $\alpha(\cdot)$, and
that $\tau_{K}=-1 \Leftrightarrow \rho_S=1 \Leftrightarrow
X_{2}=\beta(X_{1})$ for some strictly decreasing $\beta(\cdot)$
\citep{Embrechts99_176}.  Other copula based measures of pairwise
concordance exist \citep{Scarsini1984_201}, as well as multivariate
extensions \citep{Taylor2007_789}. A discussion of such measures is
beyond the scope of this paper.

Asymptotic tail dependence (hereinafter, simply tail dependence) of a random pair $(X_{1},\,X_{2})$ is another example of dependence property that is completely described by the copula and not by the marginals. The joint distribution of $(X_{1},\,X_{2})$ is said to be upper tail dependent if the probability that one of the two variables takes values in its upper tail (\ie, high quantiles), given that the other has taken values in its upper tail, does not decay to zero. Lower tail dependence is defined analogously for low quantiles. Tail dependence thus allows for simultaneous extremes, and is for instance used to model systemic risks. Formally, $(X_{1},\,X_{2})$ with marginals $F_{1}$ and $F_{2}$ are upper tail dependent if
\begin{equation}
\lim_{u\uparrow1^{-}}\P(X_{1}>F_{1}^{-1}(u)|X_{2}>F_{2}^{-1}(u))=\lambda_{u}>0,
\label{eq:upper-tail-dependence-X}
\end{equation}
and are lower tail dependent if
\begin{equation}
\lim_{u\downarrow0^{+}}\P(X_{1}<F_{1}^{-1}(u)|X_{2}<F_{2}^{-1}(u))=\lambda_{l}>0,
\label{eq:lower-tail-dependence-X}
\end{equation}
given that these limits exist; $\lambda_{u}$ and $\lambda_{l}$ are called the upper and lower tail dependence coefficients, and can be expressed in terms of the copula $C$ of $(X_1,X_2)$ by
\begin{equation}
\lambda_{u}=\lim_{u\uparrow1^{-}}\frac{1-2u+C(u,u)}{1-u}, \quad \lambda_{l}=\lim_{u\downarrow0^{+}}\frac{C(u,u)}{u}.
\label{eq:tail_dependence_U}
\end{equation}

\subsection{Copula examples}

Here we provide three families of copulas that will be used in
\secref{horizontal_truss} and \secref{dome} to model different
dependence structures among input loads on a truss model. A list of
classical families of copulas and their properties can be found in
\citet{Nelsen2006, Joe2015}. A summary of $19$ families of bivariate
copulas used for inference in this study and of their dependence
properties is provided in Tables
\ref{tab:pair_copula_cdfs}-\ref{tab:pair_copula_properties}.

The independence copula 
\begin{equation}
\CInd(\u) = \prod_{i=1}^{M}u_{i}
\label{eq:indep-copula}
\end{equation}
describes the case of mutual independence among the random variables. For $M=2$, $\CInd$ has Spearman's rho $\rho_{S}^{(\Pi)}=0$, Kendall's tau $\tau_{K}^{(\Pi)}=0$, and tail dependence coefficients $\lambda_{u}^{(\Pi)}=\lambda_{l}^{(\Pi)}=0$.

A Gaussian random vector $\X$ with correlation matrix $\corrmat=(\rho_{ij})_{i,j=1}^{M}$ and marginals $F_i \sim \N(\mu_{i},\sigma_{i}^2)$, $i=1,\ldots,M$, has copula
\begin{equation}
\Cgauss(\u)=\frac{1}{\sqrt{\det\corrmat}}\exp\left(-\frac{1}{2}\begin{pmatrix}\Phi^{-1}(u_{1})\\
\vdots\\
\Phi^{-1}(u_{M})
\end{pmatrix}^{T}\cdot\left(\corrmat^{-1} - \idmat \right)\cdot\begin{pmatrix}\Phi^{-1}(u_{1})\\
\vdots\\
\Phi^{-1}(u_{M})
\end{pmatrix}\right),
\label{eq:gaussian-copula}
\end{equation}
where $\Phi$ is the univariate standard normal $\cdf$ and $\idmat$ is the identity matrix of rank $M$. $\Cgauss$ is called Gaussian copula or normal copula. One can prove that, if $M\geq3$ variables are coupled by a Gaussian copula with correlation matrix $\corrmat,$ any pairs $(X_{i},\,X_{j})$ are coupled by a Gaussian pair copula with correlation matrix 
$\left[\begin{smallmatrix}
1 & \rho_{ij}\\
\rho_{ij} & 1
\end{smallmatrix}\right]$. 
If so, their Spearman's rho is  $\rho_S^{(\N)}=\frac{6}{\pi}\arcsin(\frac{\rho_{ij}}{2})$, their Kendall's tau is $\protect{\tau_{K}^{(\N)}=\frac{2}{\pi}\arcsin(\rho_{ij}})$, and their tail dependence coefficients are $\lambda_{u}^{(\N)}=\lambda_{l}^{(\N)}=0$.
Therefore, multivariate Gaussian copulas assign negligible probabilities to joint extremes.

A pair copula that contemplates upper tail dependence is the bivariate Gumbel-Hougaard (or Gumbel, for brevity) copula
\begin{equation}
\Cgumb(u,v)=\exp\left(-\big[(-\log u)^{\theta}+(-\log v)^{\theta}\big]^{1/\theta}\right),\;\theta\in[1,\,+\infty).
\label{eq:gumbel-hougaard-copula}
\end{equation}
In particular, if $\theta=1$ then $C^{(\mathcal{GH})}(u,v)=uv$ (the independence copula). $C^{(\mathcal{GH})}$ has Kendall's tau $\tau_{K}^{(\mathcal{GH})}=(\theta-1)/\theta$ and upper tail dependence coefficient $\lambda_{u}^{(\mathcal{GH})}=2-2^{1/\theta}$, which increases from $0$ to $1$ as $\theta$ increases from $1$ to $+\infty$. Finally, $\lambda_{l}^{(\mathcal{GH})}=0$.

\subsection{Vine copulas}
\label{subsec:Vine-Copulas}

When the input dimension $M$ grows, defining a suitable $M$-copula
which properly describes the pairwise and higher-order dependencies
among the input variables becomes increasingly difficult. Multivariate
extensions of several families of pair-copulas exist, but they rarely
fit real data well. \citet{Bedford2002} proved that, instead,
one may construct any $M$-copula by a product of simpler $2$-copulas.
Some are unconditional copulas among pairs of random variables, others
are conditioned on the values taken by other variables. Here we briefly
introduce this construction, known as pair copula or vine copula construction, and recall some important features. For details,
we refer to the cited literature (see also \cite{Czado_webpage}). A recent review with a focus on
financial applications can be found in \cite{Aas2016}.

Let $\u_{\overline{i}}$ be the vector obtained from the vector $\u$ by removing its $i$-th component, \ie, $\u_{\overline{i}} = (u_{1},\ldots,u_{i-1},u_{i+1},\ldots,u_{M})^{\tr}$. Similarly, let $\bm{u}_{\overline{\{i,j\}}}$ be the vector obtained by removing the $i$-th and $j$-th component, and so on. For a general subset $\A \subset \{1,\ldots,\,M\}$, $\u_{\notA}$ is defined analogously. Also, $\FcondA$ and $\fcondA$ indicate the joint $\cdf$ and $\pdf$ of the random vector $\X_{\notA}$ conditioned on $\X_{\A}$; $\A=\{i_{1},\ldots,i_{k}\}$ and $\notA=\{j_{1},\ldots,\,j_{l}\}$ form a partition of $\{1,\ldots,\,M\}$, that is, $\A\cup\notA=\{1,\ldots,\,M\}$ and $\A\cap\notA=\emptyset$. Using \eqrefp{f-c-rel}, $\fcondA$ can be expressed as
\begin{equation}
\begin{aligned}\fcondA(\x_{\notA}|\x_{\A})= & \ccondA(F_{j_{1}|\A}(x_{j_{1}}|\x_{\A}),\,F_{j_{2}|\A}(x_{j_{2}}|\x_{\A}),\ldots,\,F_{j_{l}|\A}(x_{j_{l}}|\x_{\A}))\times\\
 & \times\,\prod_{j\in\notA}f_{j|\A}(x_j|\x_{\A}),
\end{aligned}
\label{eq:sklar-cond-f-c-rel}
\end{equation}
where $\ccondA$ is an $l$-copula density -- that of the conditional random variables $(X_{j_{1}|\A},\,X_{j_{2}|\A},\ldots,\,X_{j_{l}|\A})^\tr$ --
and $f_{j|\A}$ is the conditional $\pdf$ of $X_{j}$ given $\X_{\A}$,
$j\in\notA$. Following \cite{Joe1996}, the univariate conditional
distributions $F_{j|\A}$ can be further expressed in terms of any
conditional pair copula $C_{ji|\A\backslash\{i\}}$ between $X_{j|\A\backslash\{i\}}$
and $X_{i|\A\backslash\{i\}}$, $i\in\A$:
\begin{equation}
F_{j|\A}(x_{j}|\x_{\A})=\frac{\partial C_{ji|A\backslash\{i\}}(u_{j},\,u_{i})}{\partial u_{i}}\big\vert_{(F_{j|\A\backslash\{i\}}(x_{j}|\x_{\A\backslash\{i\}}),\,F_{i|\A\backslash\{i\}}(x_{i}|\x_{\A\backslash\{i\}}))}.
\label{eq:Joe-condF-as-pairC}
\end{equation}

An analogous relation readily follows for conditional densities:
\begin{equation}
\begin{aligned}f_{j|\A}(x_{j}|\x_{\A})= & \frac{\partial F_{j|\A}(x_{j}|\x_{\A})}{\partial x_{j}}\\
= & c_{ji|A\backslash\{i\}}(F_{j|\A\backslash\{i\}}(x_{j}|\x_{\A\backslash\{i\}}),\,F_{i|\A\backslash\{i\}}(x_{i}|\x_{\A\backslash\{i\}}))\times\\
 & \times\,f_{j|\A\backslash\{i\}}(x_{j}|\x_{\A\backslash\{i\}}).
\end{aligned}
\label{eq:Joe-condf-as-pairc}
\end{equation}

Substituting iteratively \eqrefp{Joe-condF-as-pairC}-\eqrefp{Joe-condf-as-pairc} into \eqrefp{sklar-cond-f-c-rel}, \citet{Bedford2002} expressed $\fX$ as a product of pair copula densities multiplied by $\prod_{i}f_{i}$. Recalling \eqrefp{f-c-rel}, it readily follows that the associated joint copula density $c$ can be factorised into pair copula densities. Copulas expressed in this format are called vine copulas.

The factorisation is not unique: the pair copulas involved in the
construction depend on the variables chosen in the conditioning equations
\eqrefp{Joe-condF-as-pairC}-\eqrefp{Joe-condf-as-pairc} at each
iteration. To organise them, \citet{Bedford2002} introduced a graphical
model called the regular vine (R-vine). An R-vine among $M$ random
variables is represented by a graph consisting of $M-1$ trees $T_{1},\,T_{2},\ldots,\,T_{M-1}$,
where each tree $T_{i}$ consists of a set $N_{i}$ of nodes and a
set $E_{i}$ of edges $e=(j,k)$ between nodes $j$ and $k$. The
trees $T_{i}$ satisfy the following three conditions:
\begin{enumerate}
\item Tree $T_{1}$ has nodes $N_{1}=\{1,\text{\ensuremath{\ldots},\,M}\}$
and $M-1$ edges $E_{1}$
\item for $i=2,\ldots,\,M-1$, the nodes of $T_{i}$ are the edges of $T_{i-1}$
: $N_{i}=E_{i-1}$
\item Two edges in tree $T_{i}$ can be joined as nodes of tree $T_{i+1}$
by an edge only if they share a common node in $T_{i}$ (proximity
condition)
\end{enumerate}
To build an R-vine with nodes $\mathscr{N}=\{N_{1},\ldots,N_{M-1}\}$
and edges $\mathscr{E}=\{E_{1},\ldots,E_{M-1}\}$, one defines for
each edge $e$ linking nodes $j=j(e)$ and $k=k(e)$ in tree $T_{i}$,
the sets $I(e)$ and $D(e)$ as follows:
\begin{itemize}
\item If $e\in E_{1}$ (edge of tree $T_{1}$), then $I(e)=\{j,\, k\}$ and $D(e)=\emptyset$,
\item If $e\in E_{i}$, $i \geq 2$, then $D(e)=D(j) \cup D(k) \cup (I(j) \cap I(k))$ and $I(e) = (I(j) \cup I(k)) \backslash D(e)$.
\end{itemize}
$I(e)$ contains always two indices $j_e$ and $k_e$, while $D(e)$ contains $i-1$ indices for $e\in E_i$. One then associates each edge $e$ with the conditional pair copula $C_{j_e, k_e|D(e)}$ between $X_{j_e}$ and $X_{k_e}$ conditioned on the variables with indices in $D(e)$. An R-vine copula density with $M$ nodes can thus be expressed as \citet{Aas2016}
\begin{equation}
c(\u)=\prod_{i=1}^{M-1} \prod_{e \in E_i} c_{j_e, k_e|D(e)}(u_{j_e|D(e)},\,u_{k_e|D(e)}).
\label{eq:R-vine}
\end{equation}
 
Two special classes of R-vines are the drawable vine (D-vine,
\citep{KurowickaCooke2005_inbook}) and the canonical vine (C-vine,
\citep{Aas2009}). Denoting $F(x_{i})=u_{i}$ and
$F_{i|\A}(x_{i}|\x_{\A})=u_{i|\A}$, $i\notin\A$, a C-vine density is
given by the expression
\begin{equation}
c(\u)=\prod_{j=1}^{M-1}\prod_{i=1}^{M-j}c_{j,j+i|\{1,\ldots,j-1\}}(u_{j|\{1,\ldots,j-1\}},\,u_{j+i|\{1,\ldots,j-1\}}),
\label{eq:C-vine}
\end{equation}
while a D-vine density is expressed as
\begin{equation}
c(\u)=\prod_{j=1}^{M-1}\prod_{i=1}^{M-j}c_{i,i+j|\{i+1,\ldots,i+j-1\}}(u_{i|\{i+1,\ldots,i+j-1\}},\,u_{i+j|\{i+1,\ldots,i+j-1\}}).
\label{eq:D-vine}
\end{equation}

\begin{figure}
\begin{centering}
\includegraphics[width=8cm]{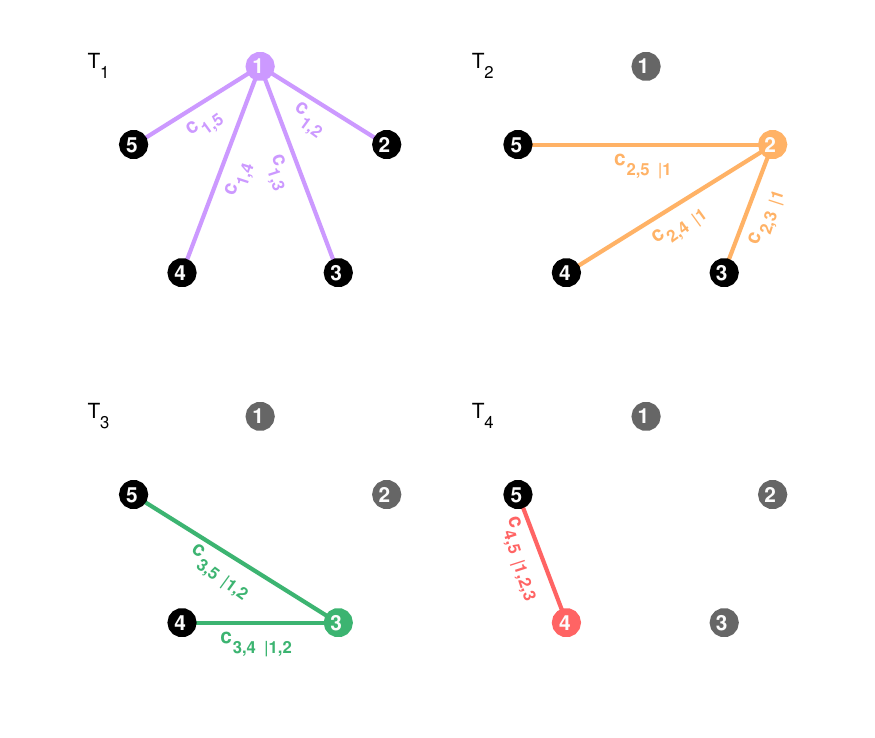}$\negthickspace$\includegraphics[width=4cm]{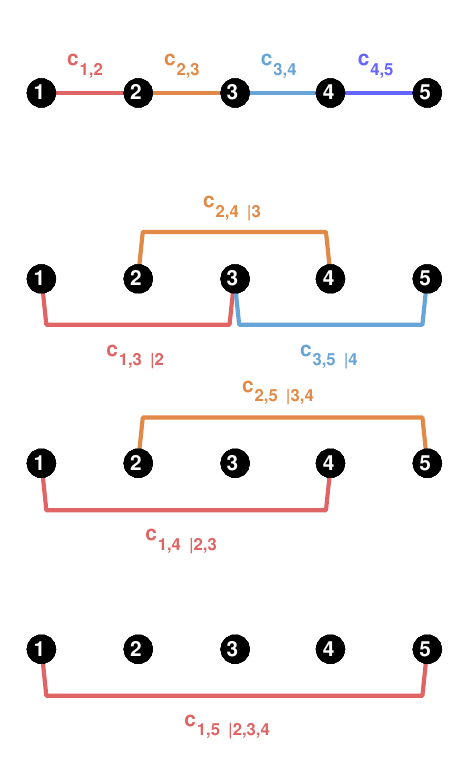}\caption{\textbf{Graphical representation of C- and D-vines.} The pair copulas in each tree of a $5$-dimensional C-vine (left; conditioning variables are shown in grey) and of a $5$-dimensional D-vine (right; conditioning variables are those between the connected nodes). 
\label{fig:vines-5dim}}
\par\end{centering}
\end{figure}

The graphs associated to a $5$-dimensional C-vine and to a $5$-dimensional D-vine are shown in \figref{vines-5dim}. Note that this simplified
illustration differs from the standard one introduced in \citet{Aas2009} and commonly used in the literature. 

\subsection{Vine inference in practice}
\label{subsec:Vine-construction-in-practice}

Building a vine copula model that properly describes the dependencies among the inputs involves the following steps:
\begin{enumerate}
\item selecting the structure of the vine (for C- and D-vines: selecting the order of the nodes);
\item modelling each pair copula in the vine by a suitable parametric family (based on expert knowledge, when available, or learning from data);
\item assigning the copula parameters (from prior knowledge, or by fitting to data).
\end{enumerate}

Steps 1-2 solve the representation problem, by providing a parametric
model of the input dependencies. Step 3 uniquely determines the copula
to be assigned to the inputs. We restrict our attention to the case
where expert knowledge is not available and the vine has to be learned
entirely from available data. \citet{Aas2009} provided algorithms to
compute the likelihood of a C- or D-vine model given a sample
$\{\hat{\x}^{(1)},\ldots,\,\hat{\x}^{(n)}\}$ of observations.
\citet{Joe2015} presented a likelihood estimation algorithm for general
R-vines. These algorithms enable, for a given parametric model (that is,
once the vine structure and comprising pair copula families have been
selected), parameter estimation (step 3) based on maximum likelihood.

The estimation could then in principle be iterated across all possible
structures (step 1) and pair copula families (step 2) to find the most
likely model describing the observed dependence properties. The number
of possibilities to loop across, however, is extremely large: an
$M$-copula density admits $2^{\binom{M-2}{2}-1}M!$ different R-vine
factorisations \citep{MoralesNapoles2011_inbook}, $M!$ of which are C-
or D- vines. This approach is thus computationally demanding in the
presence of even a moderate number of
inputs. 
In the case studies examined in this work we take a different approach,
originally proposed by \citet{Aas2009} and commonly preferred in
applications, and first solve step $1$ separately. The optimal vine
structure is found heuristically by ordering the variables $X_{i}$ such
that pairs $(X_{i},\,X_{j})$ with the strongest dependence are captured
first, \ie, fall in the first trees of the vine. The Kendall's tau
$\tau_{K; ij}$ defined in \eqrefp{Kendalls-tau-from-C} is taken as the
measure of dependence. For a C-vine, this means selecting the central
node in tree $T_{1}$ as the variable $X_{i_{1}}$ that maximises
$\sum_{j\neq i_{1}}\tau_{K; i_{1}j}$, then the node of tree $T_{2}$ as
the variable $X_{i_{2}}$ which maximises
$\sum_{j\notin\{i_{1},i_{2}\}}\tau_{K; i_{2}j}$, and so on. For a
D-vine, this means ordering the variables
$X_{i_{1}},X_{i_{2}},\ldots,\,X_{i_{M}}$ in the first tree so as to
maximise $\sum_{k=1}^{M-1}\tau_{K; i_{k}i_{k+1}}$, which we solve as an
open travelling salesman problem (OTSP) \citep{Applegate2006}. An open
source Matlab implementation of a genetic algorithm to solve the OTSP is
provided in \citet{Kirk2014_OTSP}. An algorithm to find the optimal
structure for R-vines has been proposed in \citet{Dissmann2013_52}.

Once the vine structure has been selected, steps $2$ and $3$ are solved together by an iterative procedure. For each pair copula composing the vine, and for each parametric families allowed for that copula, the parameters of the family are fitted to the available data based on maximum likelihood (other approaches, such as Bayesian estimation, may be followed \citep{Gruber2015_937}). The parametric family which best fits the data is then chosen as the family that minimizes the Akaike information criterion (AIC)
\[
\AIC=-2\log L + 2k,
\]
where $k$ is the number of parameters of the pair copula and $\log L$ is its log-likelihood. The AIC penalises models with a larger number of parameters (which typically yield higher likelihood and would otherwise be preferred), thus preventing overfitting. Alternatives to the AIC have been proposed, for instance the Bayesian information criterion (BIC) and the copula information criterion (CIC; \citep{Gronneberg2014}). Also, one may alternatively opt for various goodness of fit tests \citep{Schepsmeier2015,Fermanian2012_inbook}. We did not consider these different approaches here. For a comparison of some of them, see \citet{Manner2007_compcriteria}. Optionally, once each pair copula has been separately selected by this iterative approach (sequential fitting), the selected pair-copula families are retained and the parameters of the vine are globally fitted to the data. This step, however, may be computationally very demanding if $M$ is large.

To facilitate inference we rely on the commonly used \emph{simplifying
  assumption} that the pair copulas $C_{j(e),k(e)|D(e)}$ in
\eqrefp{R-vine} only depend on the variables with indices in $D(e)$
through the arguments $F_{i(e)|D(e)}$ and $F_{j(e)|D(e)}$
\citep{Czado2010}. While being exact only in particular cases, this
assumption is usually not severe \citep{Haff2010_simplifiedPCC}. In
\citet{Stoeber2013_101} construction techniques for non-simplified vine
copulas were proposed.

\tabref{pair_copula_cdfs} shows the list of the $19$ simplified pair
copula families used for vine copula construction in this study, and
implemented in the VineCopulaMatlab package by \citet{Kurz2014_CDVine}.
A summary of their properties is reported in
\tabref{pair_copula_properties}. In addition to these copulas, their
rotated versions were also considered. A rotation by $180^\circ$
transforms a copula into its survival version. A rotation by $90^\circ$
or $270^\circ$ implements negative dependencies. Including the rotated
copulas, $62$ families were considered in total for inference.

%
%
%

\section{Vine representations for UQ methods assuming independent inputs}
\label{sec:vines4UQ}

Some advanced UQ techniques require or benefit from inputs $\X$ with independent components. For instance, PCE (\subsecref{PCE}) exploits independence to build a basis of polynomials orthonormal with respect to $\FX$ by tensor product. This in turn simplifies the construction of a metamodel that expresses $Y$ as a polynomial of the inputs. FORM and SORM (\subsecref{FORM}), as well as other reliability methods, take advantage of the probability measure of the standard normal space to approximate low probability mass regions. When the components of $\X$ are mutually dependent but independence is needed, it is therefore custom to transform $\X$ into a vector $\Z$ with independent components. The transformation $\mathcal{T}$ that performs this mapping thus changes the copula $\CX$ of $\X$ into the independence copula $\CInd$ defined in \eqrefp{indep-copula}. When $\mathcal{T}$ also makes $F_{\Z}$ rotationally invariant, it is called an isoprobabilistic transform. This section discusses existing isoprobabilistic transformations, relates them to copula theory, and highlights the existence of algorithms for their computation when $\CX$ is expressed as an R-vine. By doing so, we demonstrate that vine copulas provide effective models of complex input dependencies also in combination with UQ approaches designed for independent inputs.

\subsection{Compositional models for dependent inputs}
\label{subsec:Compositional-models}

Consider a generic UQ method that works in a probability space where input parameters are independent and have marginal distributions $G_{i}$. Assume that the input $\X$ to the model $\M$ has a joint $\cdf$ $\FX$ for which an invertible isoprobabilistic transform $\mathcal{T}:\,\X\mapsto\Z$ is known, and that $\mathcal{T}^{-1}:\,\Z\mapsto\X$ is also known. Then, the system response $Y=\M(\X)$ can be expressed as a function of $\Z$ by

\begin{equation}
Y=(\M\circ\mathcal{T}^{-1})(\Z).
\label{eq:compositional-model}
\end{equation}

The \emph{compositional model} $\M\circ\mathcal{T}^{-1}$ can be seen
as a black box model which combines the known map $\mathcal{T}^{-1}$
with the original computational model $\M$. The UQ method of choice
can then be applied on the input $\Z$ and the model $\M\circ\mathcal{T}^{-1}$:
the statistics of the output of $\M\circ\mathcal{T}^{-1}$ in response to
$\Z$ are identical to the statistics of the output of $\M$ in response to $\X$.

Given $\FX$ and $\M$, determining the compositional model requires
then to determine $\mathcal{T}^{-1}$, which depends on $\FX$. However,
a general closed form expression for $\mathcal{T}^{-1}$ is in most
cases unknown, even when $\FX$ is known. This problem is associated exclusively to the copula $\CX$ of
$\X$, and not to its marginals $F_{i}$. Indeed, $\X$ can be mapped
by the transformation $\mathcal{T}^{(\mathcal{U})}$ defined in \eqrefp{Uniform-transformation}
onto $\bm{U}\sim U([0,\,1]^{M})$, whose joint $\cdf$ is $\CX$.
Thus, one can always write 
\begin{equation}
\mathcal{T}=\mathcal{T}^{(\mathcal{U})}\circ\mathcal{T}^{(\Pi)},
\label{eq:Isoprobabilistic-transform-composed}
\end{equation}
where $\mathcal{T}^{(\mathcal{U})}$ -- which depends on the marginals
only -- is known for a given $\FX$, while $\mathcal{T}^{(\Pi)}:\bm{\,U}\mapsto\bm{Z}$
is to be determined.

\subsection{Isoprobabilistic transforms and copulas}
\label{subsec:Rosenblatt-and-Nataf-transforms}

The most general isoprobabilistic transform $\mathcal{T}$, valid for any continuous $\FX$, is the Rosenblatt transform \citep{Rosenblatt52},
which reads
\begin{equation}
\mathcal{T}_{1}^{(\mathcal{R})}: \X \mapsto \bm{W},\,\textrm{where }\begin{cases}
W_{1} = F_{1}(X_{1})\\
W_{2} = F_{2|1}(X_{2}|X_{1})\\
\;\vdots\\
W_{M} = F_{M|1,\ldots,M-1}(X_{M}|X_{1},\ldots,\,X_{M-1})
\end{cases}.
\label{eq:Rosenblatt-transform-1}
\end{equation}

Following \eqrefp{Isoprobabilistic-transform-composed}, and as first noted in \citet{Lebrun2009c}, one can rewrite $\mathcal{T}_{1}^{(\mathcal{R})}=\TPRa\circ\TU$, where
\begin{equation}
\TPRa: \, \bm{U}\mapsto\bm{W},\, \textrm{with } W_i = C_{i|1,\ldots,i-1}(U_i|U_1,\ldots,U_{i-1}).
\label{eq:Rosenblatt-transform-Copula-1}
\end{equation}
Here, $C_{i|1,\ldots,i-1}$ are conditional copulas of $\X$ (and
therefore of $\bm{U}$), obtained from $\CX$ by differentiation.
The problem of obtaining an isoprobabilistic transform of $\X$ is
thus reduced to the problem of computing derivatives of $\CX$.

The variables $W_{i}$ are mutually independent and marginally uniformly distributed in $[0,1]$. To assign $W_{i}$ any other marginal distribution
$\Psi_{i}$, one can define the generalised Rosenblatt transform as a map $\R=\mathcal{T}_{2}^{(\mathcal{R})}\circ\mathcal{T}_{1}^{(\mathcal{R})}=\mathcal{T}_{2}^{(\mathcal{R})}\circ\TPRa\circ\TU$,
where
\begin{equation}
\mathcal{T}_{2}^{(\mathcal{R})}: \bm{W} \mapsto \Z, \, \textrm{with } Z_{i}=\Psi_{i}^{-1}(W_{i}).
\label{eq:Rosenblatt-transform-2}
\end{equation}
When $\Psi_{i}=F_{i}$ for all $i$, \ie, $\mathcal{T}_{2}^{(\mathcal{R})} \equiv \left(\TU\right)^{-1}$,
$\R$ maps $\X$ onto a random vector $\Z$ with same marginals but independent components. 

Each continuous joint $\cdf$ $\FX$ defines multiple transforms of the
type \eqrefp{Rosenblatt-transform-Copula-1}, one per permutation of the
indices $\{1,\ldots,\,M\}$. However, these transforms involve
conditional probabilities which are not generally available in closed
form. A notable exception is the multivariate Gaussian distribution,
where independence can be obtained by diagonalisation of the correlation
matrix (\eg, by Choleski decomposition). The Rosenblatt transform in
this case (and in this case only, see \citet{Lebrun2009c}) is equivalent
to the Nataf transform \citep{Nataf62}, which is commonly used in
engineering applications.

A generalized Nataf transform for elliptical copulas was proposed in \citet{Lebrun2009b}. The generalization enables the mapping of random vectors with elliptical copulas into their standard spherical representative, having uncorrelated (but not mutually independent, except for the Gaussian case) components with elliptical, unit variance marginal distributions. Adopting the generalized Nataf transform instead of the Rosenblatt transform for inputs with non-elliptical copulas, of for inputs with elliptical copulas when a transformation to independent components is needed, may cause non-negligible errors on the estimates computed by UQ methods.

\subsection{Rosenblatt transform and resampling for R-vines} \label{subsec:vine-Rosenblatt}

\citet{Aas2009} provided algorithms to compute the Rosenblatt transform
\eqrefp{Rosenblatt-transform-Copula-1} and its inverse when $\CX$ is a
given C- or D-vine. Given the pair-copulas $C_{ij}$ in the first tree of
the vine, the algorithms first compute their derivatives $C_{i|j}$.
Higher-order derivatives $C_{i|ijk}$, $C_{i|ijkh}$, $\ldots$ are
obtained from the lower-order ones and their inverses by iteration. The
derivatives of continuous pair copulas are available analytically in few
cases (see, \eg, \citet{Schepsmeier2014_525}) and numerically otherwise.
Since these functions are monotone increasing distributions, their
inverses are numerically cheap to compute by rootfinding, when not
available analytically. An algorithm for the computation of the
Rosenblatt and inverse Rosenblatt transforms for general R-vines was
proposed in \cite{Schepsmeier2015}. These algorithms can be trivially
implemented so as to process $n$ samples in parallel.

In addition, $\Rinv$ allows to sample from the vine model by
transforming independent points uniformly distributed in $[0,\,1]^{M}$.
Space filling samples in the probability space can be obtained
analogously (\eg, by Sobol sequences or Latin Hypercube sampling,
\citep{McKay1979}). Given in particular a vine model of the input
dependencies (obtained, \eg, from expert knowledge or by inference from
available data), the inverse Rosenblatt transform enables resampling
from this model.

\section{UQ for mutually dependent inputs}
\label{sec:UQ-techniques}

After recalling convergence properties of MC estimates, we summarise here three established UQ methods used in the numerical experiments carried out in Sections \ref{sec:horizontal_truss} and \ref{sec:dome}: PCE, FORM, and IS. Several other methods exist to solve the same problems, and we do not advocate for the ones considered here over others. Importantly, the framework demonstrated for these three methods extend to basically any UQ technique designed for problems with a finite number of coupled inputs.

PCE is a spectral method that expresses the system response as a polynomial of the input variables. It is used to estimate moments of the response, to compute sensitivity indices, or to perform resampling efficiently. FORM is a reliability analysis method designed to approximate small failure probabilities $P_f$ numerically. IS is a stochastic sampling method that combines FORM with MC to obtain more robust estimates of $P_f$. When the computational budget is limited and only few runs of the computational model can be afforded, these methods provide significantly better estimates of their target statistics than MCS with the same number of observations. However, these methods strongly rely on an accurate representation of the input dependencies. Besides, some of them strongly benefit from the possibility of mapping the input random vector onto a vector with independent components.

Here we describe how to combine these methods with the vine representation of the input $\cdf$ illustrated in \secref{Copulas-and-vines}. The flexibility of R-vine models expands the applicability of these methods drastically.

\subsection{Convergence of MC estimates} \label{subsec:MC_convergence}

MC (or sample) estimates of a statistic $\eta=\eta(Y)$ are obtained as functions $\hat{\eta}_n=\hat{\eta}_n(Y)$ of $n$ i.i.d realisations $\{ \hat{y}^{(j)} \}_{j=1}^{n}$ of $Y$. Three statistics considered in the applications in Sections \ref{sec:horizontal_truss}-\ref{sec:dome} are the mean $\mu(Y)$ of $Y$, its standard deviation $\sigma(Y)$, and failure probabilities of the type $P_{f; y^*}(Y) = \P(Y \geq y^{*})$, where $y^*$ is a critical threshold (see \tabref{MCestimates}, first row). Their sample estimators are the sample mean $\hat{\mu}_{n}(Y)$, the corrected sample standard deviation $\hat{\sigma}_{n}(Y)$, and the sample survival function evaluated at $y^*$, $\hat{P}_{f;y^*, n}(Y)$. Their analytical expression is given in \tabref{MCestimates}, second row.
	
If $\hat{\eta}_n(Y)$ is an unbiased estimator of $\eta(Y)$ and $\eta(Y) \neq 0$, the reliability of $\hat{\eta}_n(Y)$ can be quantified by its coefficient of variation (CoV), given by
\[
\textrm{CoV}(\hat{\eta}_n(Y))=\frac{\sigma(\hat{\eta}_n(Y))}{\mu(\hat{\eta}_n(Y))}=\frac{\sigma(\hat{\eta}_n(Y))}{\eta(Y)}.
\]

\begin{table}
	\begin{center}
	\begin{tabular}{rccc}
		\toprule
		 $\eta(\cdot)$:  & $\mu(Y)=\int_{\RR} y f_Y(y)dy$  
		                 & $\sigma(Y)=\sqrt{\int_{\RR}(y-\mu(Y))^2 f_Y(y)dy}$
		                 & $P_{f;y^*}(Y)=\int_{\{ Y \geq y^* \}} f_Y(y) dy$ \\[1em]		   
 		 $\hat{\eta}_n(\cdot)$:  & $\displaystyle{\hat{\mu}_n(Y)=\frac{1}{n}\sum_j \hat{y}^{(j)}}$   
                         & $\displaystyle{\hat{\sigma}_n(Y)=\sqrt{\frac{1}{n-1}\sum_j \left(\hat{y}^{(j)}-\hat{\mu}_n(Y)\right)^2}}$ 
                         & $\displaystyle{\hat{P}_{f;y^*}(Y)=\frac{1}{n}\sum_j \mathbf{1}_{\{ \hat{y}^{(j)}>y^* \}}}$ \\[1.5em]
		 $\textrm{CoV}{(\hat{\eta}_n)}$: & $\displaystyle{\frac{\sigma(Y)}{\mu(Y)} \frac{1}{\sqrt{n}}}$ 
						 & $\displaystyle{\approx \frac{1}{\sqrt{2n}}}$ 
						 & $\displaystyle{\sqrt{\frac{1-P_f}{n P_f}}}$ \\
		 \bottomrule
 	\end{tabular}
	\end{center}
 	\caption{\textbf{Some MC sample estimates and their CoV}. The first row of the table defines the mean, standard deviation, and failure probability of the random variable $Y$. The second row shows their sample estimators, and the bottom row the CoV of such estimators (exact for $\hat{\sigma}_n(Y)$ only if $Y$ is normally distributed).}
 	\label{tab:MCestimates}
\end{table}

It is common in engineering applications to accept estimates whose CoV
is not larger than $0.1$ ($10\%$).  The CoV of all statistics in
\tabref{MCestimates}, third row (approximate for $\hat{\sigma}_n(Y)$) is
proportional to $1/\sqrt{n}$ and thus decays to $0$ as $n$ increases,
however at a slow pace. The expression for
$\textrm{CoV}(\hat{\sigma}_{n}(Y))$ is obtained from the fact that
$\sigma(\hat{\sigma}_{Y}) = \sigma(Y)/\sqrt{2N +O(N^2)}$ if $Y$ is
normally distributed (see \citet{Romanovski1925}).

\subsection{Polynomial Chaos Expansion}
\label{subsec:PCE}

PCE is a spectral method that represents a model $\M$ of finite variance as a linear sum of orthogonal polynomials \citep{Ghanembook2003, Xiu2002}. As such, the parameters of the resulting representation have a statistical interpretation. For instance, the first two moments of the PCE model are encoded in the coefficients of the obtained polynomial. The model is also computationally cheap to evaluate, enabling an efficient evaluation of other global statistics of $Y$ (higher order moments, the $\pdf$, \emph{etc.}) that would otherwise require an excessive number of runs of $\M$.

Building a PCE representation of the output is relatively simple, as recalled below, for independent inputs. For this reason, if $\X$ has non mutually independent components, it is convenient to first map it onto such a vector $\Z$ by an isoprobabilistic transform. Modelling the copula $\CX$ of $\X$ as an R-vine provides the Rosenblatt transform \eqrefp{Rosenblatt-transform-Copula-1}-\eqrefp{Rosenblatt-transform-2} needed to this end.

Given an isoprobabilistic transform $\mathcal{T}$ such that $\Z=\mathcal{T}(\X)$, it follows that $Y=(\M\circ\mathcal{T}^{-1})(\Z)=\M^{\prime}(\Z)$. In the following, $Y$ is assumed to have finite variance. The PCE of $Y=\M^{\prime}(\Z)$ is defined as
\begin{equation}
Y=\sum_{\bm{\alpha} \in \mathbb{N}^{M}} y_{\alpha} \Psialpha(\Z),
\label{eq:PCE-def}
\end{equation}
where the $\Psi_{\bm{\alpha}}$ are multivariate polynomials orthonormal with respect to $f_{\Z}$, \ie,
\[
\int_{\D_{\Z}}\Psialpha(\z)\Psibeta(\z)f_{\Z}(\z)d\z=\delta_{\bm{\alpha\beta}}.
\]
Here, $\delta_{\bm{\alpha\beta}}$ is the Kroenecker delta symbol.

Since $\Z$ has independent components, each $\Psialpha$ can be obtained as a tensor product of $M$ univariate polynomials $\phialpha(x_{i})$ orthonormal with respect to the marginals $g_{i}$ of $Z_i$:
\[
\Psialpha(\z)=\prod_{i=1}^{M}\phialpha(z_{i}).
\]

The polynomial basis is guaranteed to exist if the marginals
distributions all have finite moments of any order. A unique
representation exists if additionally the marginals are uniquely
represented by the sequence of their moments. For details, as well as
for sufficient conditions that guarantee uniqueness, see
\cite{Ernst2012_317}. For instance, the $\phialpha$ are Hermite
polynomials if $Z_{i}$ is standard normal, \ie, if
$g_{i}(z)=\varphi(z)=\exp(-z^{2}/2)/\sqrt{2\pi}$. In the applications
illustrated in \secref{horizontal_truss} we work with this choice,
although other choices may be favoured in different applications. An
investigation of optimal choices for the marginal distributions of $\Z$
is an open question that will be investigated in a future study.
Classical families of polynomials are described in
\citet{Xiu2002}. 

The sum in \eqrefp{PCE-def} comprises an infinite number of terms. For practical purposes, it is truncated to a finite sum \citep{UQdoc_10_104}. Given a truncation scheme and the corresponding set $\A$ of multi-indices, the coefficients $y_{\bm{\alpha}}$ in 
\begin{equation}
Y_{PC}(\Z)=\sum_{\bm{\alpha}\text{\ensuremath{\in\A}\ }}y_{\alpha}\Psialpha(\Z)
\label{eq:PCE-truncated}
\end{equation}
are evaluated on a set
$\{(\hat{\z}^{(j)}=\mathcal{T}^{-1}(\hat{\x}^{(j)}),\,\hat{y}^{(j)}\}_{j=1}^{n}$
of observations (the \emph{experimental design}). Many strategies exist
to accomplish this task, such as projection methods based on Gaussian
\citep{Lemaitre01} or sparse quadrature \citep{Keese03, XiuBook2010},
least-squares minimisation \citep{Berveiller2006a}, and different
adaptive sparse methods \citep{Doostan2011, Jakeman2015}, hybridised into
a single methodology by \citep{Blatman2011_JCP}. The latter is
particularly suitable when $M$ is large, because it achieves a sparse
basis out of a very large initial set of possible polynomials, and is
therefore the method of choice in this study.

Once a PCE metamodel \eqrefp{PCE-truncated} of the compositional model $\M^{\prime}$ is built, the first two moments of the response are encoded in the coefficients of the expansions. Indeed, due to orthonormality of the polynomial basis,
\begin{equation}
\E(Y_{PC})=y_{\bm{0}}
\label{eq:PCE-mean-var}, \quad \V(Y_{PC})=\sum_{\alpha\in\A\backslash\{\bm{0}\}}y_{\bm{\alpha}}^{2}.
\end{equation}

Higher-order moments, as well as other statistics, can be efficiently estimated by simulation.

\subsection{First order reliability method}
\label{subsec:FORM}

Let $Y=\M(\X)$ be the uncertain, scalar output of the computational model $\M$ in response to an uncertain $M$-variate input $\X$ with joint $\cdf$ $\FX$, joint $\pdf$ $\fX$ and domain $\DX$. Suppose that the system fails if $Y \geq y^{*}$, where $y^{*}$ is a critical threshold. The failure condition is usually rewritten as $g(\x)\leq0$, with $g(\x)=y^{*}-\M(\x)$. 

Reliability analysis concerns the evaluation of the failure probability $P_{f}=\P(g(\X) \leq 0)$, \ie, of the probability mass over the failure domain $\D_{f}=\lbrace\omega:\;g(\X(\omega))\leq0\rbrace$. $\D_f$ is typically known only implicitly, preventing a direct estimation of $P_{f}$. 

Using the indicator function 
\[
\bm{1}_{\D_{f}}(\x)=\begin{cases}
1 & \textrm{if }g(\x)\leq0\\
0 & \textrm{if }g(\x)>0
\end{cases},
\]
$P_{f}$ can be expressed as 
\begin{equation}
P_{f}=\int_{\DX}\bm{1}_{\D_{f}}(\x)\fX(\x)d\x=\E(\bm{1}_{\D_{f}}(\X)),
\label{eq:Pf_as_mean}
\end{equation}
where $\E(\text{\ensuremath{\cdot}})$ is the mean operator with respect to $\fX$. 

If $\X$ is multivariate normal with independent components, FORM \citep{Hasofer1974, Hohenbichler1983} approximates $\D_f$ with an hyperplane tangent to the limit-state surface $\lbrace\omega:\;g(\X(\omega))=0\rbrace$ in its point closest to the origin (the \emph{design point} $\x^*$). The rationale is that the standard normal density is a fast decaying function of the norm of its argument, so that -- assuming the uniqueness of the design point -- the probability mass of $\D_f$ concentrates around $\x*$ (see also \cite{DK86}).

If $\X$ is not multivariate normal, but has a normal copula, the Nataf
transform (which is equivalent to the Rosenblatt transform in the
Gaussian case, see \citet{Lebrun2009c} and \subsecref{vine-Rosenblatt})
is used map $\X$ into a standard normal random vector
$\Z=\mathcal{T}(\X)$, and FORM can then be used to search for the design
point $\z^*$ in the standard normal space. If $\X$ has a more general
elliptical copula, the generalized Nataf transform can be employed to
map $\X$ into a vector $\Z$ whose components are uncorrelated (but not
independent) and have elliptical marginals with unit variance
\citep{Lebrun2009b}. In this case, the probability density of $\Z$ is
again a rapidly decreasing function of the norm of its argument, and
FORM can be used analogously to the standard normal case.

If $\X$ has a non-elliptical copula, employing the generalized Nataf transform would yield biased estimates of the failure probability. One has to resort to different isoprobabilistic transformations, the most general being the Rosenblatt transform, to map $\X$ into a standard normal (or into a spherical elliptical) random vector $\Z=\mathcal{T}(\X)$, thus reconducting the problem to one of the two cases above. To treat this more general case, we express $\FX$ in terms of its marginals $F_{i}$ and of its copula $\CX$ as in \eqrefp{Sklar-F-C-relation}, and we model $\CX$ as an R-vine (see \subsecref{Vine-Copulas}). The Rosenblatt transform of the latter is available (see \subsecref{vine-Rosenblatt}), allowing $\X$ to be mapped onto the standard normal space, where the classical version of FORM can be used.

\subsection{Importance sampling} \label{subsec:IS}

In the context of reliability analysis, IS is used to combine the convergence speed of FORM with the robustness of MC sampling. For a general, non-standard-normal input vector $\X$ admitting Rosenblatt transform $\Z=\T(\X)$, \eqrefp{Pf_as_mean} can be recast as 
\begin{equation}
P_f = \int_{\RRM} \bm{1}_{\D_{f}} \left( \mathcal{T}^{-1}(\z) \right) \frac{\varphi_M(\z)}{\psi(\z)} \psi(\z) d\z   
    = \E_{\psi} \left( \bm{1}_{\D_{f}} \left( \mathcal{T}^{-1}(\Z) \right) \frac{\varphi_M(\Z)}{\psi(\Z)} \right),
\end{equation}
where $\varphi_M(\cdot)$ is the M-variate standard normal density,
$\psi(\cdot)$ is a suitable M-variate density (the \emph{importance
  density}) and $\E_{\psi}$ is the mean operator with respect to $\psi$.
\citet{Melchers1999} recommends to assign $\psi$ as the standard normal
density centered at the design point found by FORM:
$\psi(\z)=\varphi_M(\z-\z^*)$.

Given a sample $\{ \hat{\z}^{(1)}, \,\ldots, \hat{\z}^{(n)} \}$ of $\psi(\Z)$, the IS estimator of $P_f$ is the sample estimator
\[
\PfIS = \frac{1}{N} \exp(||\z^*||^2/2) \sum_{j=1}^{n} \bm{1}_{\D_{f}} \left( \mathcal{T}^{-1}(\hat{\z}^{(j)}) \right) \exp(-\hat{\z}^{(j)} \cdot \z^*),
\]
which has variance 
\[
\V(\PfIS) \approx \frac{1}{n(n-1)}\sum_{j=1}^n \left( \bm{1}_{\D_{f}} \left( \mathcal{T}^{-1}(\hat{\z}^{(j)}) \right) \frac{\varphi_M(\hat{\z}^{(j)})}{\varphi_M(\hat{\z}^{(j)}-\z^*)} -\PfIS \right)^2
\]
and $\textrm{CoV}(\PfIS) \approx \std(\PfIS)/\PfIS$. Since the latter is given in terms of $\PfIS$, it is unknown until $\PfIS$ is computed. One can progressively increase the sample size $n$ until $\textrm{CoV}(\PfIS)$ drops below a desired level.

\section{Results on a horizontal truss model}
\label{sec:horizontal_truss}

We first demonstrated the analysis workflow developed above on a
horizontal truss model. Estimates based on advanced and computationally
efficient UQ techniques were compared to reference MCS estimates.
Earlier work on vine representations combined with MCS can be found in
\cite{Wang2017_inpress}, limited to reliability analysis.

The analysis was run in Matlab \citep{Matlab2016a}. Specifically, the
vine copula inference was performed using the open source package
VineCopulaMatlab \citep{Kurz2014_CDVine}. We enriched this toolbox with
functionalities for the computation of the Rosenblatt and inverse
Rosenblatt transforms of C- and D- vines, and for the calculation of the
optimal D-vine structure on data, implemented as an open travelling
salesman problem \citep{Applegate2006}. The UQ analyses were performed
with the free Matlab-based software $\uqlab$
\citep{Marelli2014}.

\subsection{Computational model}

\begin{figure}
	\begin{center}
		\includegraphics[width=10cm]{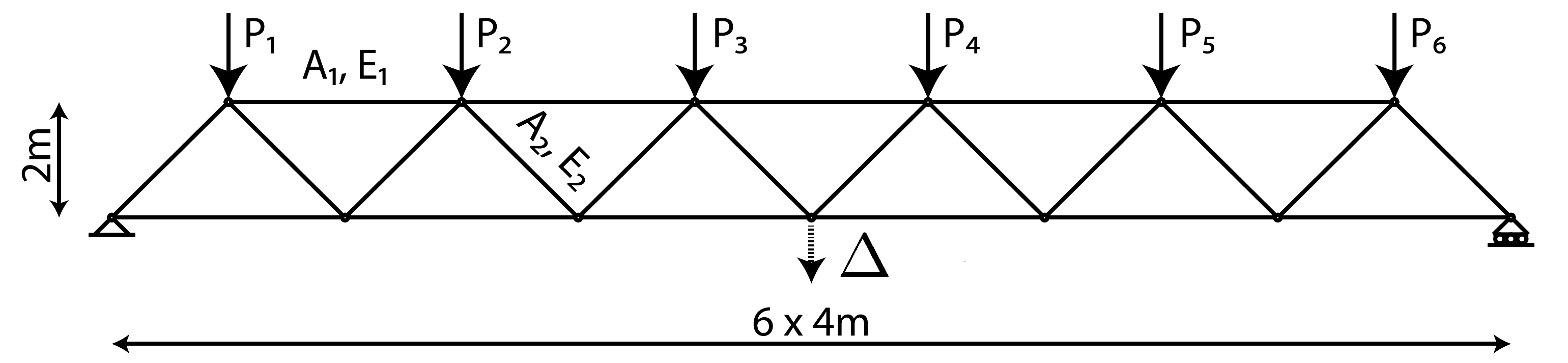}
		\caption{\textbf{Scheme of the horizontal truss model.} Modified from \citet{Blatman2011_JCP}.}
		\label{fig:Truss_model_Blatman}	
	\end{center}
\end{figure}

The horizontal truss model, already used in \citet{Blatman2011_JCP}, comprises $23$ bars connected at $6$ upper nodes, as shown in \figref{Truss_model_Blatman}. The structure is $24$ meters long and $2$ meters high. Six random loads $P_{1},\,P_{2},\,\ldots,\,P_{6}$ were applied onto the structure, one on each upper node. As a result, the structure exhibited a downward vertical displacement at each node. The largest displacement $\Delta$ was always at the center. Excess displacement leads to failure. 

\subsection{Probabilistic input model}

We considered the case of uncertain loads $P_i$, causing uncertainty in the output response $\Delta$. The bar properties of the truss, differently from \citet{Blatman2011_JCP}, were kept constant. The loads $\X=(P_{1},\,P_{2},\,\ldots,\,P_{6})$ were modelled by assigning separately their marginals $F_{i}$ and their copula $\CX$. We fixed the marginals to univariate Gumbel $\cdf$s with mean $\mu=5\times10^{4}\Pa$ and standard deviation $\sigma=0.15\mu=7.5\times10^{3}\Pa$:
\begin{equation}
F_{i}(x;\alpha,\beta)=e^{-e^{-(x-\alpha)/\beta}},\quad x \in \RR, \,i=1,\,2,\ldots,6,
\label{eq:Gumbel-marginals}
\end{equation}
where $\beta=\sqrt{6}\sigma/\pi$, $\alpha=\mu-\gamma\beta$, and $\gamma\approx0.5772$ is the Euler-Mascharoni constant.

We then investigated how different copulas affect the statistics of the truss response. First, we employed the independence copula $\CInd$ defined by \eqrefp{indep-copula}, which implies independence among the loads. 

Loads on a truss structure may be expected to be positively correlated: higher loads on one node increase the chance to have higher loads on other nodes (e.g. due to snow or traffic jam on a bridge). To account for this, we selected next a $6$-dimensional Gaussian copula $\CGauss$ \eqrefp{gaussian-copula}. We assigned the copula parameters $\rho_{1j}$, $j=2,\ldots,6$, such that the Spearman's correlation coefficients \eqrefp{Spearman-rho-from-C} would be $\rho_{S; 1j}=0.135$, resulting in $\rho_{1j}=0.141$ (and Kendall's tau $\tau_{K; 1j} \simeq 0.0904$). Besides, we set $\rho_{S;ij|1}=0$ for each $i\neq j$, $i,j\neq1$, so that all loads other than $P_1$ would be conditionally independent given $P_{1}$. 

Beside being positively correlated, in a realistic scenario loads are likely to be upper tail dependent: an extremely large load on one node increases the chance to have large loads elsewhere (\eg, when due to a heavy snowfall or to a traffic jam). Therefore, we last investigated a scenario with tail dependent loads (see \subsecref{Copula-measures-of-dep}). We modelled upper tail dependence by means of a C-vine $\CVine$. We selected $P_{1}$ as the first node of $\CVine$, and we set the pair copulas in the first tree to  bivariate Gumbel-Hougaard copulas $C_{1j}^{(\mathcal{GH})}$ \eqrefp{gumbel-hougaard-copula} between $P_{1}$ and $P_{j}$, $j=2,\ldots,\,5$. We took the parameter $\theta_{1j}=1.1$, $j=2,\ldots,6$, yielding Spearman's correlation coefficients $\rho_{S; 1j}=0.135$ as for the Gaussian copula. This choice resulted in $\tau_{K;1j}=0.0909$ (close to the value determined by the Gaussian copula) but also determined an upper tail dependence coefficient $\lambda_{u;1,j}=0.122$ between $P_{1}$ and $P_{j}$. We further set the other pair copulas of the vine, \ie, all conditional pair copulas, to the independence copula, ensuring conditional independence of $(P_{i},P_{j})$ given $P_1$ for each $i,\,j\neq1$. The resulting vine $\CVine$ had density
\begin{equation}
\cVine(u_{1},\ldots,u_{6})=\prod_{j=2}^{6}c_{1j}^{(\mathcal{GH})}(u_{1},u_{j}).
\label{eq:TrussVine}
\end{equation}

Note that other vine structures could have been used to model tail dependent loads: for instance, a D-vine whose first tree couples the loads on neighbouring nodes of the truss. Expert knowledge and available input data may provide guidance in this selection process.

We finally investigated the viability of the vine representation when the vine itself is not known and has to be fully inferred from data. To this end, we sampled $m=300$ realisations from $\CVine$ and learned from them the vine structure, its pair copula families and their parameters, as detailed in \subsecref{Vine-construction-in-practice}. The pair copulas were chosen among the parametric families listed in \tabref{pair_copula_cdfs} and their rotated versions defined by \eqrefp{rotated_copulas}. The inferred pair copulas comprising $\CVineHat$ are summarised in \tabref{pair_copulas_inferred_horiztruss_vinetrue}, along with their Kendall's tau and upper tail dependence coefficients. In real applications, the input observations needed for the inference procedure may be obtained by monitoring of the loads themselves, or may be estimated from available data (\eg, weather or traffic conditions), and do not require any model evaluation. The resulting C-vine $\CVineHat$ had a different structure, only two of the five pair copulas in the first tree were of the Gumbel-Hougaard family, and one of the conditional copulas in the second tree was not the independence copula. All other pair copulas were correctly found to be independence copulas. Despite the differences from the true vine $\CVine$, using $\CVineHat$ provided very good quality estimates of the statistics of the truss deflection, as shown below.

\begin{table}
	\begin{center}
		\begin{tabular}{lllll}
			\toprule
			Copula & Family & Parameter values & $\tau_K$ & $\lambda_u$ \\
			\cmidrule{1-5}
			$C_{13}$ & Clayton, rotated 180  & $\theta = 0.1806$                                                 & $0.0828$ & $0.0215$ \\ 
			$C_{12}$ & Tawn-2 & $\theta_1=1.439$, $\theta_2=0.3406$                                              & $0.1522$ & $0.1975$ \\ 
			$C_{16}$ & Gumbel & $\theta=1.103$                                                                   & $0.0934$ & $0.1254$ \\
			$C_{14}$ & Tawn, rotated 180 & $\theta_1=7.506$, $\theta_2=0.05197$, $\theta_3=0.2982$ \hspace{12pt} & $0.0454$ & $0$  \\
			$C_{15}$ & Clayton, rotated 180 & $\theta = 0.3794$                                                  & $0.1595$ & $0.1609$     \\
			$C_{35|1}$ & Tawn & $\theta_1=11.05$, $\theta_2=0.1338$, $\theta_3=0.1178$                           & $0.0658$ & $0.1151$ \\
			\bottomrule
		\end{tabular}
		\caption{\textbf{Pair copulas of inferred C-vine for loads on horizontal truss.} Pair copulas of the C-vine model $\CVineHat$ for the loads on the horizontal truss model, obtained from $300$ samples of $\CVine$. The pair copulas found to be independence copulas are not shown. The last two columns indicate the Kendall's tau and the upper tail dependence coefficient of each pair copula.}
		\label{tab:pair_copulas_inferred_horiztruss_vinetrue}
	\end{center}
\end{table}

\subsection{Analysis of the moments for different load couplings}

For each probabilistic model of the loads, we analysed the mean $\mu(\Delta)$ and standard deviation $\sigma(\Delta)$ of the resulting system response by MCS and by PCE.

The MC estimates were computed as sample estimates on $\lbrace\hat{\delta}_{i}=\M(\hat{\x}_{i})\rbrace_{i=1}^{10^7}$, where $\lbrace\hat{\x}_{i}\rbrace_{i=1}^{10^7}$ was a set of i.i.d input realisations
. The vertical deflections $\delta_i$ were computationally affordable to compute due to the simplicity of the model. The results are summarised in \tabref{TrussResponse-Moments-MC}, together with the CoV of the estimates (see \tabref{MCestimates}).
 
While $\mu(\Delta)$ was virtually identical across the input model, $\sigma(\Delta)$ exhibited non-negligible changes. For instance, it increased by almost $10\%$ from the independence to the vine copula. 
As a consequence, if $\CVine$ were the true copula among the loads, an MC estimate of $\sigma(\Delta)$ based on the independence assumption would be biased. Conversely, fitting a Gaussian copula or a C-vine to data yielded more accurate estimates. 

\begin{figure}
	\centering{}
	\includegraphics[width=16cm]{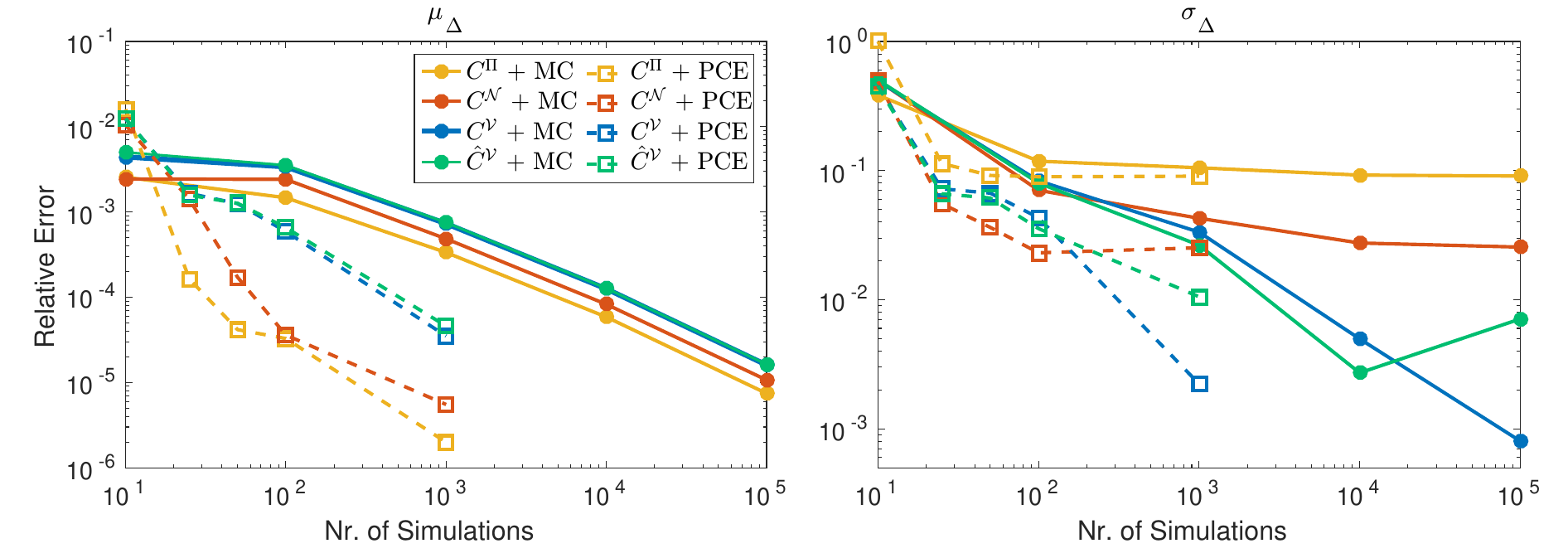}
	\caption{\textbf{Estimates of the deflection moments.} Estimates of the errors on $\mu(\Delta)$ (left panel) and on $\sigma(\Delta)$ (right panel) obtained using an increasing number of samples by MCS (solid lines) and by PCE (dashed lines), for loads coupled by $\CInd$ (yellow), $\CGauss$ (red), $\CVine$ (blue) and $\CVineHat$ (green). Reference solution: MC estimate obtained on $10^7$ samples with copula $\CVine$.}
	\label{fig:truss_PCEresults}
\end{figure}

MC estimates converge slowly (see \subsecref{MC_convergence}) and therefore need to be computed on large samples. \figref{truss_PCEresults} shows by solid lines the errors on MC estimates drawn for the four different copulas on $10,\,100,\,\ldots,10^5$ samples. The reference solutions were the MC estimates $\muVineMC$, $\sigmaVineMC$ obtained through $10^7$ samples under $\CVine$. The errors of estimates $\tilde{\mu}$, $\tilde{\sigma}$ were defined as
\begin{equation}
E_{\textrm{rel}}^{(\mu)}=\abs{\frac{\tilde{\mu}}{\muVineMC}-1},\quad E_{\textrm{rel}}^{(\sigma)}=\abs{\frac{\tilde{\sigma}}{\sigmaVineMC}-1}.
\label{eq:RelativeError}
\end{equation}
Note that, due to the CoV of the reference solutions, reported in \tabref{TrussResponse-Moments-MC}, the errors shown in \figref{truss_PCEresults} are reliable only down to approximately $10^{-4}$ for the means and $10^{-3}$ for the standard deviations.

\begin{table}
	\begin{center}
		\begin{tabular}{rcccc}
			\toprule
			& $\CInd$ & $\CGauss$ & $\CVine$ & $\CVineHat$\\
			\cmidrule{2-5}
			$\muMC(\Delta)$ (cm): & $\muIndMC=7.78$ & $\muGaussMC=7.78$ &
			                        $\muVineMC=\bm{7.78}$ & $\muVineHatMC=7.78$ \\
			$\textrm{CoV}(\muMC(\Delta))$ ($\times 10^{-5}$): & $2.1$ & $2.3$ & $2.4$ & $2.4$ \\[.5em]
			$\sigmaMC(\Delta)$ (cm): & $\sigmaIndMC=0.528$ & $\sigmaGaussMC=0.566$ &
			                        $\sigmaVineMC=\bm{0.581}$ & $\sigmaVineHatMC=0.593$ \\
  			$\textrm{CoV}(\sigmaMC(\Delta))$ ($\times 10^{-4}$): & $2.2$ & $2.2$ & $2.2$ &  $2.2$ \\
  			\bottomrule 
		\end{tabular}		
	\end{center}
	\caption{\textbf{Moments of horizontal truss deflection for different load couplings.} MC estimates $\muMC(\Delta)$ and $\sigmaMC(\Delta)$ for different copulas of the loads, based on $10^7$ samples, and their CoV. Reference solutions in bold.}
	\label{tab:TrussResponse-Moments-MC}
\end{table}

We further estimated for each copula the error on $\mu(\Delta)$ and $\sigma(\Delta)$ yielded by PCE, which is known to converge faster than MCS (see \subsecref{PCE}). We increased the size of the experimental design from $10$ to $1000$ sample points. The errors on the PCE estimates are shown in \figref{truss_PCEresults} by dashed lines (again, reliable only down to the above mentioned precision). Notably, for the same number $n$ of samples the PCE error is significantly smaller than the MCS error, demonstrating that the vine representation is fully compatible with PCE metamodelling. 

\subsection{Reliability analysis for different load couplings}

The truss model was set to fail if the deflection $\Delta$ reached or exceeded the critical threshold $\delta^{*} = 11\cm$. Reliability analysis was performed to evaluate the failure probability $P_{f}=\P(\Delta \geq \delta^{*})=1-F_{\Delta}(\delta^*)$.

For each probabilistic input model (i.e. for each copula $\CX$, combined with the marginals in \eqrefp{Gumbel-marginals}), we first obtained reference solutions by MCS. Using the $n=10^7$ i.i.d. realisations $\lbrace\hat{\delta}_{i}=\M(\hat{\x}_{i})\rbrace_{i=1}^{10^7}$ obtained for the analysis of the moments, we estimated $P_f$ as the fraction of observed deflections $\hat{\delta}_{i}$ larger than $11\cm$. Then, we drew estimates by FORM, applied on the compositional model resulting from decoupling the loads via Rosenblatt transformation (see Sections \ref{sec:vines4UQ} and \ref{subsec:FORM}). The results are summarised in \tabref{Pf-estimates}.

\begin{table}[t]
	\begin{center}				
		\begin{tabular}{rccccccccc}
			\toprule 
			Method: & \multicolumn{4}{c}{MCS} & & \multicolumn{4}{c}{FORM}\tabularnewline
			\cmidrule{2-5} \cmidrule{7-10}
			Copula:
			                      & $\CInd$         & $\CGauss$       & $\CVine$                & $\CVineHat$     & \hspace{10pt} 
			                      & $\CInd$         & $\CGauss$       & $\CVine$                & $\CVineHat$     \\
  $\hat{P}_f\,\,(\times10^{-4})$: & $0.15 \pm 0.01$ & $0.34 \pm 0.02$ & $\bm{5.04}\pm\bm{0.07}$ & $3.30 \pm 0.06$ &     
                                  & $0.037$         & $0.10$          & $4.88$                  & $2.94$ \\
  $\textrm{CoV}(\hat{P}_f)\,\,(\%)$: & $8.2$        &  $5.4$          & $1.4$                   & $2.3$           & & ---   & ---   & ---   & --- \\     
                     $\#$ runs:   & $10^7$          & $10^7$          & $10^7$                  & $10^7$          & & $219$ & $219$ & $108$ & $128$ \\
			\bottomrule
		\end{tabular}
	\end{center}
	\caption{\textbf{Estimates of the truss failure probability.} Estimates of $P_{f}$ obtained with different copulas and methods (for MCS with standard deviation of the estimator; reference solution in bold), CoV of the MC estimate (see \tabref{MCestimates}), and number of runs of the computational model needed to obtain the solution. 
	\label{tab:Pf-estimates}}
\end{table}

\begin{figure}[t]
	\begin{center}
		\includegraphics[width=8cm]{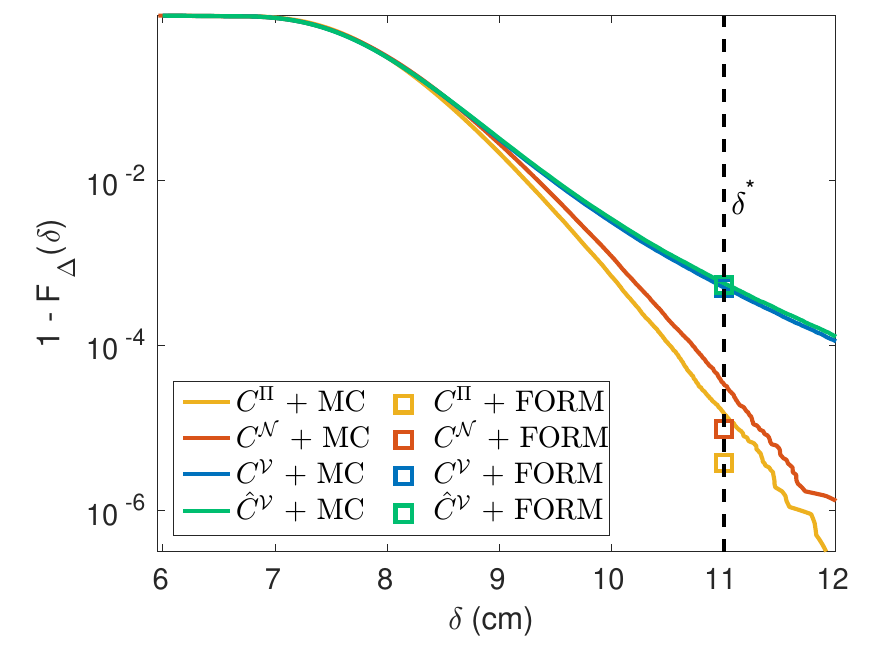}
		\caption{\textbf{Reliability analysis of the truss structure.} Solid lines: MC estimate of the survival function $\P(\Delta > \delta)$ for loads coupled by the independence copula $\CInd$ (yellow), the Gaussian copula $\CGauss$ (red), the C-vine with known parameters $\CVine$ (blue, mostly overlapping with green), and the C-vine with fitted parameters $\CVineHat$ (green). The vertical dashed line marks the critical threshold $\delta^{*}$. Squares: estimates of $P_f$ obtained by FORM.}
		\label{fig:Truss-reliability}
	\end{center}	
\end{figure}

The failure probability estimated by MCS with the independence copula $\CInd$ was $\hat{P}_{f}^{(\Pi)}=(1.5\pm0.1)\times10^{-5}$. The FORM estimate was $\PfIndFORM=0.37 \times 10^-5$, obtained by $219$ runs of the computational model. 
\figref{Truss-reliability}B shows, in yellow, the empirical survival function of $\Delta$ obtained under $\CInd$, for values of $\delta$ ranging from $6\cm$ to $12\cm$. The vertical dashed line marks the critical threshold $\delta^{*}$, whereas the square indicates the FORM estimate of $P_f$.

The MC estimate of $P_f$ under the Gaussian copula $\CGauss$ was $\PfGaussMC=(3.4 \pm 0.2)\times10^{-5}$ (\figref{Truss-reliability}, orange line: empirical survival function of $\Delta$ under $\CGauss$). Compared to the independent case, the failure probability increased by a factor of over $2$, as a result of the positive correlations among the loads. The FORM estimate was $\PfGaussFORM=1.0\times 10^{-5}$, obtained again by $219$ runs. 

The MC estimate of $P_f$ under the C-vine $\CVine$ was $\PfVineMC= (5.04 \pm 0.07)\times10^{-4}$ (\figref{Truss-reliability}, blue line: empirical survival function of $\Delta$ under $\CVine$). This value was over $33$ times larger than the case of independent loads and $14$ times larger than the Gaussian case, despite the marginal distributions of the loads being identical across all cases, and the Spearman's correlation coefficients being identical between $\CVine$ and $\CGauss$. The FORM estimate using $\CVine$ was $\PfVineFORM=4.88\times10^{-4}$, obtained by only $108$ runs. 

Finally, the MC estimate of $P_f$ assuming $\CVineHat$ was $\PfVineHatMC=3.30\times10^{-4}$, about $35\%$ smaller than the reference solution $\PfVineMC$ (\figref{Truss-reliability}, green line: empirical survival function of $\Delta$ under $\CVineHat$). The FORM estimate, obtained with $128$ runs of the computational model, was $\PfVineHatFORM=2.94\times10^{-4}$ ($42\%$ smaller than $\PfVineMC$).

In light of these results, in a scenario where the true dependence among the loads is described by \eqrefp{TrussVine}, assuming independence or a Gaussian copula would cause a severe underestimation of the failure probability of the system, even when relying on a large MCS strategy. Properly capturing the dependencies (in particular, the tail dependencies) among the inputs is thus more critical towards getting accurate estimates of $P_f$ than using more precise estimation algorithms (FORM combined with $\CVine$ outperforms MCS on $10^7$ samples combined with $\CGauss$). Furthermore, the error on $P_f$ remains small when the vine is entirely inferred from available input data, because the tail dependencies are properly captured (see \tabref{pair_copulas_inferred_horiztruss_vinetrue}). This demonstrates the viability of the vine copula modelling framework in reliability analysis for purely data driven inference. 

The results above show that the failure probability is heavily misestimated when inputs coupled by a C-vine with tail dependencies are modelled by a Gaussian copula. A natural question that arises is the following: is the opposite also true? In other words, how well can the C-vine family capture input dependencies described by a Gaussian copula? To answer this question, we performed additional simulations with $\CGauss$ as the true input copula (having associated failure probability $\PfGaussMC=(3.4 \pm 0.2)\times10^{-5}$). We sampled $300$ observations from $\CGauss$, and inferred a C-vine from those. The pair copula families were selected as before by AIC among the parametric families listed in \tabref{pair_copula_cdfs} and their rotated version, and their parameters were fitted by maximum likelihood. The resulting estimate of $P_f$ was $(2.2 \pm 0.2)\times10^{-5}$, only $35\%$ smaller than the reference value. This demonstrates that the C-vine family is an effective dependence model also in the presence of simple Gaussian dependencies. In conclusion, this class of models covers a larger range of dependence scenarios than Gaussian (or elliptical) copulas, enabling UQ also in cases where the classical use of the Nataf transform would lead to wrong estimates.

\begin{table}
	\begin{center}
		\begin{tabular}{llllc}
			\toprule
			Copula & Family & Parameter values & $\tau_K$ & $\lambda_u$ \\
			\cmidrule{1-5}
			$C_{12}$ & Partial Frank & $\theta = 1.363$                                        & $0.1189$ & $0$ \\ 
			$C_{13}$ & Partial Frank & $\theta = 1.320$                                        & $0.1166$ & $0$ \\ 
			$C_{14}$ & IterFGM       & $\theta_1=0.8832$, $\theta_2=-0.8688$ \hspace{12pt}     & $0.1536$ & $0$ \\
			$C_{15}$ & AMH           & $\theta = 0.4297$                                       & $0.1080$  & $0$  \\
			$C_{16}$ & Gaussian      & $\theta = 0.1726$                                       & $0.1104$ & $0$     \\
			\bottomrule
		\end{tabular}
		\caption{\textbf{Pair copulas of inferred C-vine for horizontal truss, under Gaussian assumption.} Pair copulas of the C-vine model for the loads on the horizontal truss model, obtained from $300$ samples of $\CGauss$. The pair copulas found to be independence copulas are not shown. The last two columns indicate the Kendall's tau and the upper tail dependence coefficient of each pair copula.}
		\label{tab:pair_copulas_inferred_horiztruss_gaussiantrue}
	\end{center}
\end{table}

\section{Results on a dome truss model under asymmetric loads} \label{sec:dome}

We further considered a more complex model of a three-dimensional $120$-bar dome truss. The model was used to demonstrate the applicability of our novel copula-based UQ framework on a more realistic case study than the one previously analysed. Due to the computational complexity of this model ($\sim 15$ seconds/run), large MCS was not affordable.

\subsection{Computational model}

\begin{figure}
	\includegraphics[width=16cm]{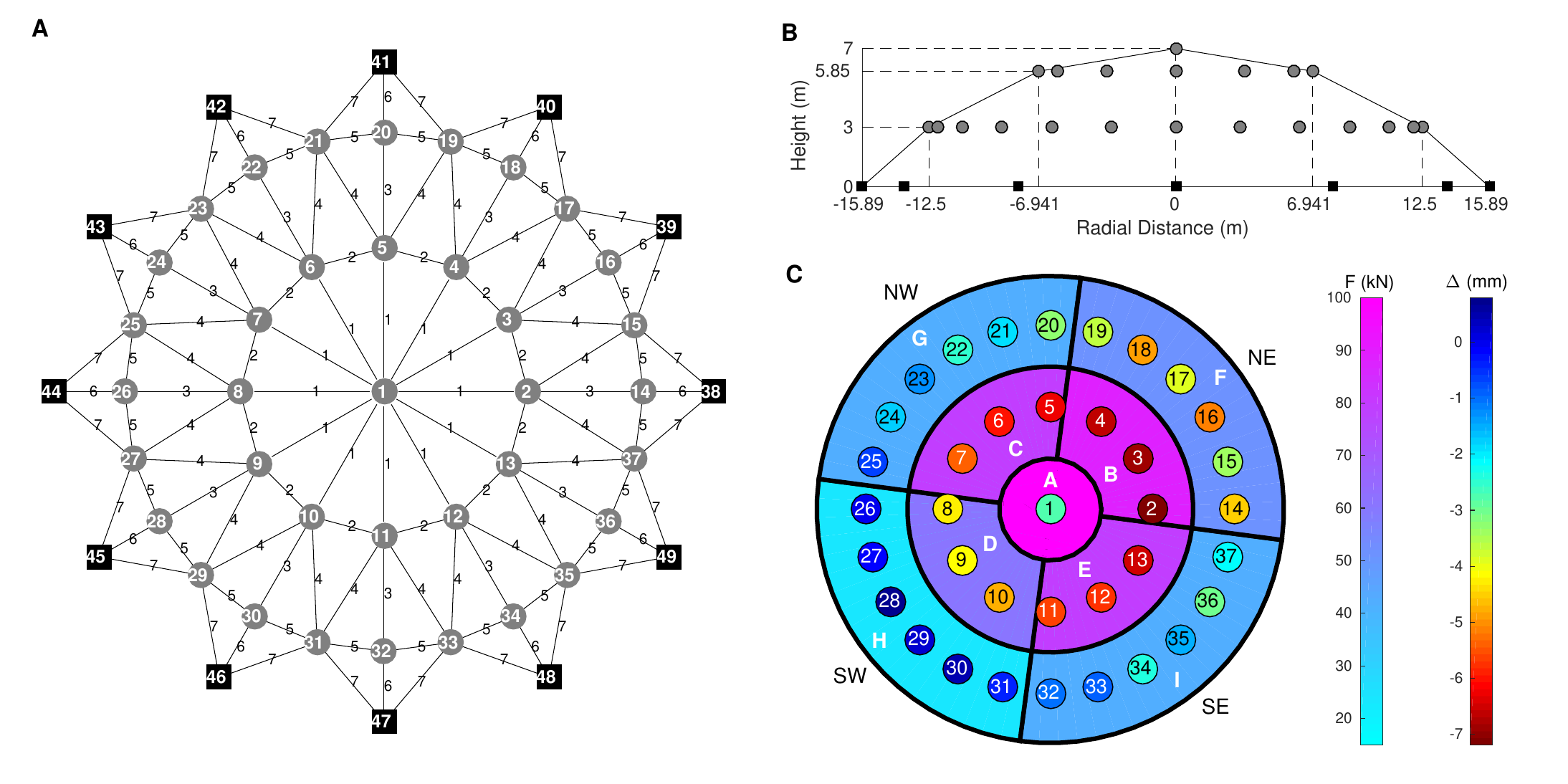}
	\caption{\textbf{Dome structure and average response to loads.} \textbf{A)} Truss model, consisting of $37$ central nodes (grey dots, $1$-$37$), $12$ support nodes (black squares, $37$-$49$) and $120$ bars connecting them, divided into $7$ groups. \textbf{B)} Profile of the dome with spatial dimensions. \textbf{C)} Sectors of the dome surface and nodes in each sector. The color of each sector represents the total average load weighing on each node in the sector (left color bar). The color of each node marks the average vertical displacement of that node in response to the loading, calculated over $1000$ Monte Carlo simulations.}
	\label{fig:dome}
\end{figure}

The dome structure, illustrated in \figref{dome}A from the top and in \figref{dome}B from the front, consists of $120$ bars connected to a total of $49$ nodes. Nodes $1$ to $37$ (grey dots) are unsupported and therefore, when subject to vertical loading, exhibit a displacement from the original position in possibly all directions. The spatial dimensions of the structure are reported in panel B. 

The computational model was implemented in the finite element software Abaqus \citep{Abaqus_v6p9}. This structure was previously analysed in \citet{Kaveh2009_267} to obtain optimal sizing variables so as to minimise the total structural weight. The authors distinguished $7$ groups of bars, and optimised the cross-sections of each group to minimise the total structural weight of the structure under $4$ different types of stress and displacement constraints. We considered in particular their case $2$, where stress and displacement constraints ($\pm 5\mm$) in the $x-$ and $y$- directions were enforced. In \citet{Kaveh2009_267} it was further assumed that each unsupported node is subject to vertical loading, taken as $60\kN$ at node 1, $30\kN$ at nodes $2$-$14$, and $10\kN$ at nodes $15-37$.  Under these conditions, the optimal cross sections reported in \tabref{dome-optimal-crosssec} were obtained, yielding a total structural weight of $89,35\kN$.

{
\begin{table}[t]
	\begin{center}
		\begin{tabular}{*9ccccc}  
		\toprule
		\multicolumn{8}{c}{Optimal cross-sectional areas (cm$^2$)} & Weight ($\mathrm{kN}$) & \multicolumn{4}{c}{Loads ($\mathrm{kN}$)} \\
		\cmidrule{2-13} 
		 Group: & 1 &      2   &    3  &   4   &   5   &   6   &   7   &   \multirow{2}{*}{89.35}  & Node: & 1 & 2-13 & 14-37  \\
		\cmidrule{2-8} \cmidrule{10-13}
		Area: & 24.38 & 21.79 & 26.61 & 17.64 & 10.38 & 22.79 & 16.38 &  & Load: & 60 & 30 & 10 \\
		\bottomrule
	\end{tabular}
	\end{center}
        \caption{\textbf{Dome's structural parameters and loads}. Cross-section of each group of bars that minimises the total structural weight under stress and $x$-, $y$- displacement constraints, resulting structural weight, and additional loads on each node. Values provided by \citet{Kaveh2009_267}.}
\label{tab:dome-optimal-crosssec}
\end{table}
}

\subsection{Probabilistic input model} \label{subsec:dome-input-uncertainties}

Here we were interested in analysing the displacement of the nodes, considered as a risk factor potentially leading to failure. 

First, we assigned uncertainty to the bar cross-sections. We modelled the $7$ previously identified groups by independent log-normal random variables with mean given by the values in \tabref{dome-optimal-crosssec} and CoV $\sigma/\mu=0.03$. We assigned the bars in each of the $7$  groups identical cross-sections.

We further assigned uncertainty to the loads applied on each node. We
modelled a scenario where loads are distributed asymmetrically over the
structure. A preliminary analysis showed that asymmetric loads yielded
higher maximum vertical displacement compared to symmetric loads. We
divided the $37$ central nodes into $9$ groups, named A to I,
corresponding to different sectors of the surface of the dome. Sector A
contains solely node $1$, sector $B$ contains nodes $2$, $3$ and $4$,
and so on, as shown in \figref{dome} and in the upper two rows of
\tabref{dome-stochastic-loads}. We considered the nodes in each group to
be subject to the same load, and modelled the loads on the $9$ groups by
a $9$-dimensional random vector with prescribed marginals and copula.
The marginals were taken to be Gumbel distributions
\eqrefp{Gumbel-marginals}, whose moments, shown in the bottom two rows
of \tabref{dome-stochastic-loads}, were determined as follows. We
assigned to each sector a Gumbel-distributed load, having mean
$1\kN/\m^2$ for the top and north-east sectors A,B,F, $0.5\kN/\m^2$ for
the north-west and south-east sectors C, E, G and I, and $0.25\kN/\m^2$
for the south-west sectors D and H. The different mean values could
model, for instance, snow falling on the dome from the north-east
direction. The average external weight on each node (third-last row of
the table) was obtained by multiplication with the total area of the
node's sector (fourth row) and by division with the number of nodes in
that sector. The CoV of each distribution was set to $0.2$. Finally,
deterministic service loads similar to those suggested by
\citet{Kaveh2009_267} were added: $60\kN$ on node $1$, $30\kN$ on nodes
2-13, $10\kN$ on nodes $14-37$.

{
\begin{table}
	\begin{center}
		\begin{tabular}{rl*9c}  
			\toprule
			Sector & {} & A &  B  &  C  &  D   &  E    &   F   &   G   &   H   &   I   \\
			Nodes & {} & 1 & 2--4 & 5--7 & 8--10 & 11--13 & 14--19 & 20--25 & 26--31 & 32--37 \\[1em]
			$\Lfix$/Node & ($\textrm{kN}$) & 60 & \multicolumn{4}{|c|}{$30$ each} & \multicolumn{4}{c}{$10$ each} \\[1em]
			Sector's Area & ($\textrm{m}^2$) & 37.84 & \multicolumn{4}{|c|}{$257.00$ each} & \multicolumn{4}{c}{$496.38$ each} \\
			Area/Node  & ($\textrm{m}^2$) & 37.84 & \multicolumn{4}{|c|}{{ }$86.33$ each} & \multicolumn{4}{c}{{ }$82.73$ each} \\
			$\Lext$/$\textrm{m}^2$ & ($\textrm{kN}$) & 1 & 1 & 0.5 & 0.25 & 0.5 & 1 & 0.5 & 0.25 & 0.5\\
			$\E(\Lext/\textrm{Node})$  & ($\textrm{kN}$) & 37.84 & 21.58 & 10.79 & 5.40  & 10.79 & 20.68 & 10.34 & 5.17 & 10.34 \\[1em]
			$\E(\Ltot/\textrm{Node})$ & ($\textrm{kN}$) & 97.84 & 51.58 & 40.79 & 35.40 & 40.79 & 30.68 & 20.34 & 15.17 & 20.34 \\
			$\std(\Ltot/\textrm{Node})$  & ($\textrm{kN}$) & 7.57 & 4.32 &  2.16 & 1.08 & 2.16 & 4.14 & 2.07 & 1.03 & 2.07 \\
			\bottomrule
		\end{tabular}
	\end{center}
	\caption{\textbf{Load statistics on each dome sector}. For each node sector from A to I: nodes in the sector, structural load $\Lfix$ per node , average external load $\Lext$ per node, moments of the total load $\Ltot$.}
	\label{tab:dome-stochastic-loads}
\end{table}
}

We coupled the $9$ loads by three different copulas: the independence copula $\CInd$ \eqrefp{indep-copula}, the Gaussian copula $\CGauss$ \eqrefp{gaussian-copula}, and a $9$-dimensional C-vine $\CVine$ \eqrefp{TrussVine}. The C-vine consisted of Gumbel-Hougaard pair-copulas $\eqrefp{gumbel-hougaard-copula}$, each with parameter $\theta=5$, between sector $A$ and sectors $B,\ldots,I$ for the first tree, and independence conditional pair-copulas for the other trees. This choice assigns the loads between nodes in sector $A$ and any other loads a Kendall's correlation coefficient $\tau_K=0.8$ and an upper tail dependence coefficient $\lambda_u=0.85$. Thus, $C^{\mathcal{V}}$ assigns a strong positive correlation to the loads and a high probability of having joint extremes if one of the loads takes values in its upper tail. The Gaussian copula was taken such that its correlation matrix would match the correlation coefficients determined by the C-Vine.

We further inferred a C-vine $\CVineHat$ from $300$ samples obtained
from $\CVine$. The resulting vine $\CVineHat$, whose comprising pair
copulas are listed in \tabref{pair_copulas_inferred_dome_vinetrue}, had
the same structure as $\CVine$, Gumbel-Hougaard copulas
$C_{AB},\,C_{AC},\,\ldots,\,C_{AH}$, and Tawn-2 copula $C_{AI}$. The
Tawn-2 copula is a generalization of the Gumbel copula with right-skewed
asymmetry in relation to the main diagonal. It is obtained from the
three-parameters Tawn copula \citep{Tawn1988_bevt} by setting one of its
two asymmetry parameters to $1$ (see Tables
\ref{tab:pair_copula_cdfs}-\ref{tab:pair_copula_properties}, row 17).
All conditional copulas of $\CVineHat$, finally, were correctly found to
be independence copulas.

\begin{table}
	\begin{center}
		\begin{tabular}{lllll}
			\toprule
			Copula & Family & Parameter values & $\tau_K$ & $\lambda_u$ \\
			\cmidrule{1-5}
			$C_{AB}$ & Gumbel & $\theta=5.093$                        & $0.8037$ & $0.8542$ \\ 
			$C_{AC}$ & Gumbel & $\theta=4.836$                        & $0.7932$ & $0.8459$ \\ 
			$C_{AD}$ & Gumbel & $\theta=5.151$                        & $0.8059$ & $0.8560$ \\
			$C_{AE}$ & Gumbel & $\theta=4.775$                        & $0.7906$ & $0.8438$ \\
			$C_{AF}$ & Gumbel & $\theta=4.631$                        & $0.7841$ & $0.8385$ \\
			$C_{AG}$ & Gumbel & $\theta=5.018$                        & $0.8007$ & $0.8519$ \\
			$C_{AH}$ & Gumbel & $\theta=4.712$                        & $0.7878$ & $0.8415$ \\
			$C_{AI}$ & Tawn-2 & $\theta_1=5.257$, $\theta_2=0.967$    & $0.7875$ & $0.8445$ \\
			\bottomrule
		\end{tabular}
		\caption{\textbf{Pair copulas of inferred C-vine for loads on dome structure.} Pair copulas of the C-vine model $\CVineHat$ for the loads on the dome structure, obtained from $300$ samples of $\CVine$. The pair copulas found to be independence copulas are not shown. The last two columns indicate the Kendall's tau and the upper tail dependence coefficient of each pair copula.}
		\label{tab:pair_copulas_inferred_dome_vinetrue}
	\end{center}
\end{table}

\subsection{Model response to the uncertain input} \label{subsec:dome-response}

The output of the model in response to a single instance of the input is a list of displacements in the $x$-, $y$- and $z$- directions, one per node, as well as the tension (or compression) of each bar. We restricted our attention to displacements only, which, if excessive ($11\mm$ in any direction), lead to failure of the structure.

We first performed a preliminary Monte-Carlo analysis based on $1000$ simulations of the input (with loads coupled by $C^{\mathcal{V}}$) and corresponding output displacements. \figref{dome}C shows in two different color codes the average weights on the nodes in each sector (left color bar) and the resulting average vertical displacement of each node (right color bar). Negative displacement indicates that the node moved downwards. Some nodes exhibited positive displacements, \ie, uplifting. 

For all simulations and all nodes, the vertical displacement always exceeded in absolute value the displacement in the $x$- and $y$- directions. This was expected, considering that the average bar's cross-sections were optimised to minimise the latter two. Besides, the absolute vertical displacement was always maximal at node $2$, except for $17$ out of $1000$ simulations where the maximal absolute displacement was observed at node $3$, but was never critical (that is, was always $<11\mm$). Thus, we reduced the model's response to the vertical displacement $\Delta$ of node $2$:
\[
\Delta=\M(\X), \quad \X=(A_1,\ldots,\,A_7, L_A,\ldots,\,L_I).
\]

\subsection{Analysis of the moments} \label{subsec:dome_PCE}

For each copula mentioned above, we evaluated the mean $\mu(\Delta)$ and the standard deviation $\sigma(\Delta)$ of the deflection $\Delta$ at node $2$ both by MCS and by PCE. The estimates were based on samples of size $n$ increasing from $10$ to $1000$. Due to the generally faster convergence of PCE with respect to MCS for small sample sizes, the PCE estimates built on $1000$ samples were taken as reference values for each of the four copula models (see \tabref{DomeResponse_Moments_PCE}). The values obtained indicate that the independence, Gaussian and vine copulas yielded for $\Delta$ similar means but different standard deviations.

\begin{table}
	\begin{center}
		\begin{tabular}{rcccc}
			\toprule
			& $\CInd$ & $\CGauss$ & $\CVine$ & $\CVineHat$\\
			\cmidrule{2-5}
			$\muPCE(\Delta)$ (mm): & $\muIndPCE=-7.193$ & $\muGaussPCE=-7.183$ & $\muVinePCE=\bm{-7.182}$ & $\muVineHatPCE=-7.182$ \\[1em]
			$\sigmaPCE(\Delta)$ (mm): & $\sigmaIndPCE=\;\;\,1.164$ & $\sigmaGaussPCE=\;\;\,0.588$ & $\sigmaVinePCE=\bm{\;\;\,0.552}$ & $\sigmaVineHatPCE=\;\;\,0.560$ \\
			\bottomrule 
		\end{tabular}		
	\end{center}
	\caption{\textbf{Moments of dome's deflection for different load couplings.} PCE estimates of $\protect\mu(\Delta)$ and $\protect\sigma(\Delta)$ for different copulas among the loads, based on $1000$ observations. Reference solutions in bold.}
	\label{tab:DomeResponse_Moments_PCE}
\end{table}

\begin{figure}[t]
	\includegraphics[width=16cm]{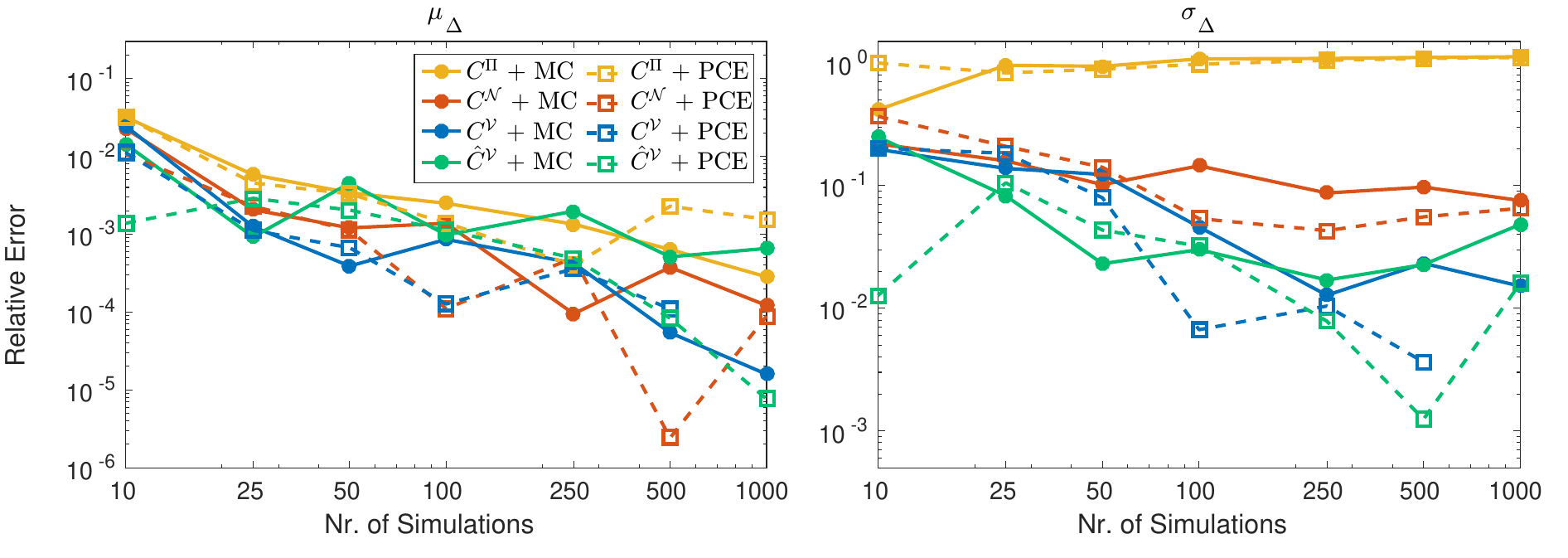}
	\caption{\textbf{Errors on moments of dome's deflection} $\bm \Delta$. Estimates of the errors on $\mu(\Delta)$ (left panel) and on $\sigma(\Delta)$ (right panel) obtained using an increasing number of samples by MCS (solid lines) and by PCE (dashed lines), for loads coupled by $\CInd$ (yellow), $\CGauss$ (red), $\CVine$ (blue) and $\CVineHat$ (green). Reference solutions: PCE estimates $\tilde{\mu}^{(\Vine)}$, $\tilde{\sigma}^{(\Vine)}$ obtained on $1000$ samples with copula $\CVine$.}
	\label{fig:dome_PCEresults}
\end{figure}

Taken in particular $\CVine$ to be the true copula among the loads, and the corresponding PCE estimates $\muVinePCE$ and $\sigmaVinePCE$ based on $1000$ points to be the reference solutions, we computed the relative error of all other estimates. The errors, defined analogously to \eqrefp{RelativeError}, are shown in \figref{dome_PCEresults}. From these results, three main conclusions can be drawn. First, if $\CVine$ was the true copula among the loads, neither $\CInd$ nor $\CGauss$ would offer adequate alternative representations, since the associated standard deviations differ significantly (by $111\%$ and $6.5\%$, respectively) from their reference value.  Second, by employing the inferred vine $\CVineHat$ in combination with MCS (green solid line) it is possible to approximate the moments with higher precision than by the Gaussian (red) or independence (yellow) copulas. Finally, PCE combines well with the vine representation (dashed lines), yielding the smallest errors. It is worth noting, however, that using a proper copula model (a vine instead of a Gaussian or independence copula) is more important to obtain accurate estimates (particularly for $\sigma(\Delta)$) than using a more advanced UQ method (PCE instead of MC). 

\subsection{Reliability analysis} \label{subsec:dome_FORM}

The dome structure was further set to fail if the displacement $\Delta$ was equal to or lower than the critical threshold $\delta^*=-11\mm$. We performed reliability analysis to estimate the failure probability $P_f=\P(\Delta \leq \delta^*)=F_{\Delta}(\delta^*)$ of excessive downward vertical displacement.

\begin{figure}
	\centering
	\includegraphics[width=8cm]{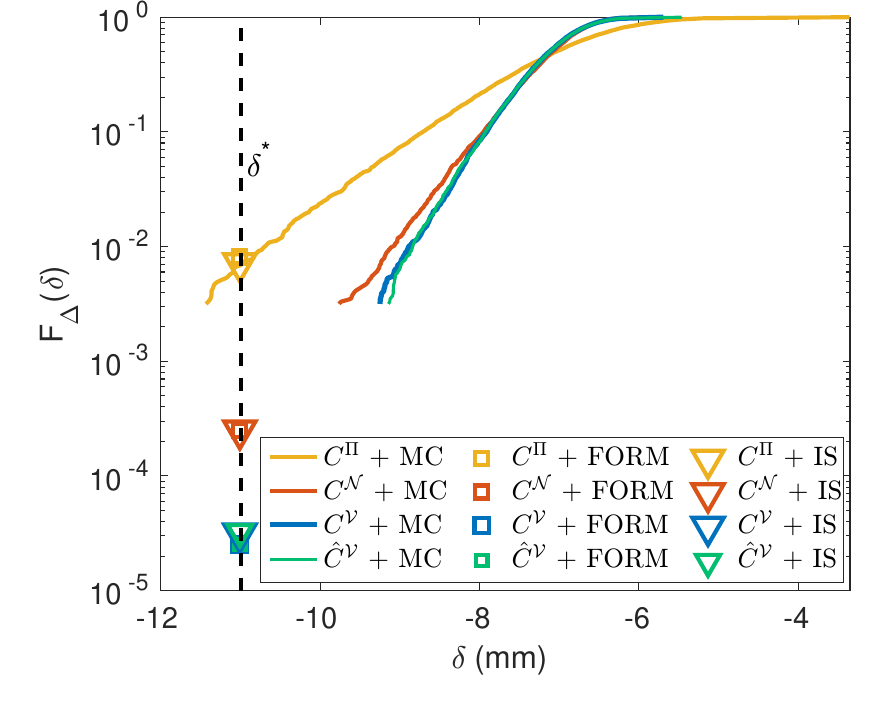}
	\caption{\textbf{Dome failure probability.} The solid lines show the MC estimates of the response $\cdf$ $F_{\Delta}(\delta)=\protect\P(\Delta \leq \delta)$ (for values down to $10^{-3}$) under loads coupled by $\CInd$ (yellow), $\CGauss$ (red), $\CVine$ (blue) and $\CVineHat$ (green). The vertical dashed line represents the critical threshold $\delta^{*}$. The squares and triangles on it indicate the FORM and IS estimates of the failure probability, respectively, obtained for each copula model.}
	\label{fig:dome_FORM}
\end{figure}

We performed $5000$ simulations by MCS for each copula of the input model, keeping the marginals identical across the models. \figref{dome_FORM} shows the $\cdf$s resulting from copulas $\CVine$ (blue), $\CVineHat$ (green), $\CGauss$ (red), and $\CInd$ (yellow), evaluated for probabilities down to $10^{-3}$.

The MC estimate of $P_f$ under $\CInd$ was $\PfIndMC=(6.8\pm 1.2)\times 10^{-3}$. For copulas $\CGauss$, $\CVine$ and $\CVineHat$ no simulations led to values of $\Delta$ below $\delta^*$. We then resorted to FORM to evaluate $P_f$ for each copula, obtaining the estimates $\PfIndFORM=8.0\times 10^{-3}$, $\PfGaussFORM=2.50\times 10^{-4}$, $\PfVineFORM=2.53\times 10^{-5}$, and $\PfVineHatFORM=2.54\times 10^{-5}$. Since MC estimates were not available or not reliable here in light of the small sample being available, we further performed IS to improve the FORM estimates and to get confidence intervals (see \subsecref{IS}). We increased the IS sample size in steps of $100$, until the CoV of the estimate was lower than $10\%$. We obtained the estimates $\PfIndIS=(7.13 \pm 0.70)\times 10^{-3}$, $\PfGaussIS=(2.47 \pm 0.24)\times 10^{-4}$, $\PfVineIS=(3.13 \pm 0.29)\times 10^{-5}$, and $\PfVineHatIS=(3.27 \pm 0.30)\times 10^{-5}$.

\begin{table}
	\begin{center}
		\resizebox{\textwidth}{!}{
		\begin{tabular}{rcccccccccccccc}
			\toprule
			Method: & \multicolumn{4}{c}{MCS} & \hspace{0pt} & \multicolumn{4}{c}{FORM} & \hspace{0pt} & \multicolumn{4}{c}{IS}\\
			\cmidrule{2-5} \cmidrule{7-10} \cmidrule{12-15}
			Copula: & $\CInd$ & $\CGauss$ & $\CVine$ & $\CVineHat$ & 
			        & $\CInd$ & $\CGauss$ & $\CVine$ & $\CVineHat$ & 
			        & $\CInd$ & $\CGauss$ & $\CVine$ & $\CVineHat$\\
    $\hat{P}_{f}\,(\times10^{-5})$: & $680\pm 120$ & -- & -- & -- & 
                    & $800$   &    $25$   & $2.53$   & $2.56$      & 
                    & $713 \pm 70$  & $24.7 \pm 2.4$  & $\bm{3.13 \pm 0.29}$ & $3.27 \pm 0.30$ \\
    $\textrm{CoV}(\hat{P}_f)$ (\%): & $17.6$ &  -- & -- & -- & 
	                & --      &   --   & --   & -- &
	                & $9.9$   & $9.9$     &  $9.3$   & $9.3$ \\     
		 tot. $\#$ runs: & $5000$  & $5000$    &  $5000$  & $5000$   & 
		            & $182$   & $349$     &  $276$   & $181$   &
		            & $482$   & $749$     &  $876$   & $781$ \\
			\bottomrule
		\end{tabular}}
		\caption{\textbf{Estimates of the dome failure probability.} Estimates of $P_{f}$ obtained with different copulas and methods (reference solution in bold), CoV of the MC and IS estimates, and number of runs needed for the estimation. 
			\label{tab:dome_Pf_estimates}}
	\end{center}
\end{table}

The results, summarized in \tabref{dome_Pf_estimates}, show that the failure probability of the structure decreases by an order of magnitude from $\CInd$ to $\CGauss$, and by another order of magnitude from $\CGauss$ to $\CVine$ and $\CVineHat$. Highly asymmetric loads (as due to $\CInd$ and, to a minor extent, to $\CGauss$) may create a deformation mechanism in the structure that favours large displacements of the most heavily loaded nodes (here, node $2$). In contrast, the more symmetric loading determined by the C-vine results in a more evenly distributed load path that ultimately leads to a safer structure. For loads actually coupled by $\CVine$, assuming the independence or Gaussian copulas thus leads to highly overestimating $P_f$. Conversely, building the vine by purely data-driven inference recovers the reference solution $\PfVineIS$ with high precision. Again, the input model used for the analysis is more important to get an accurate estimate than the particular UQ method (FORM or IS) employed.

\section{Discussion} \label{sec:discussion}

We proposed a general framework that enables uncertainty quantification (UQ) for problems where the input parameters of the system exhibit complex, non-Gaussian, non-elliptical dependencies (copulas). The joint $\cdf$ of input parameters is expressed in terms of marginals and a copula, which are modelled separately. The copula is further modelled as a vine copula, \ie, a product of simpler $2$-copulas. This specification eases its construction, especially in high dimension, and offers a simple interpretation of the dependence model. A wide range of different dependence structures can be modelled using this approach.

Our framework focuses in particular on regular (R-) vines, for which
algorithms exist to compute the likelihood on available data, thus
enabling parameter fitting and data driven inference. In addition,
R-vines offer algorithms to compute the associated Rosenblatt transform
and its inverse on data, used to map the original input random vector
into a vector with independent components and back. Thus, UQ techniques
that benefit from input independence can be applied to any inputs
coupled by R-vines. In this work we restricted our attention to inputs
with continuous marginals, which cover a large class of engineering
problems. Extensions of R-vines to discrete (\eg, categorical and count)
data have been recently proposed \citep{Panagiotelis2012_1063,
  Panagiotelis2017_138}.

The methodology was first demonstrated on a simple horizontal truss model, for which Monte Carlo solutions were computationally affordable, and then replicated on a more complex truss model of a dome. Both structures deflected in response to loads on different nodes. Changing the copula among the loads from the independence to a Gaussian to a tail-dependent C-vine copula changed the statistics of the deflection, in particular its variance and upper quantiles. Taken the vine copula as the true dependence structure among the loads, the independence and Gaussian assumptions thus led to biased estimates of these statistics. The failure probability of the two systems, in particular, was mis-estimated by one to two orders of magnitude. This was true regardless of the particular UQ method used for the estimation. Using instead a vine copula model of the input dependencies and fitting the model to relatively few input observations yielded far better estimates. 

These results demonstrate that using a proper dependence model for the inputs can be more critical to get high-accuracy estimates of the output statistics than employing a superior UQ algorithm. Our framework encompasses both aspects, allowing highly flexible probabilistic models of the input to be combined with virtually any UQ technique designed to solve problems characterized by (finitely many) coupled inputs. Also, we demonstrated that a suitable vine representation can be properly inferred on data also in the presence of simple Gaussian dependencies. Thus, this class of dependence models effectively covers a broader range of problems than the Nataf transform (also in its generalized form by \citet{Lebrun2009b}) does.

Selecting a vine that properly represents the dependencies of
multivariate inputs may be challenging. We discussed and employed
existing methods to perform fully automated inference on available data.
When the dimension of the input is large or the parametric families of
pair copulas considered for the vine construction are many, this
approach may become computationally prohibitive. A-priori information on
the input statistics may be used to ease the selection, for instance by
Bayesian methods \citep{Gruber2015_937}. The problem of selecting
suitable vines, however, remains open in very high dimension (say,
$>50$) or on very large samples. Also, computing the Rosenblatt and
inverse Rosenblatt transforms in these cases may be computationally
demanding or lead to numerical instability. Separating the inputs into
mutually independent subgroups, by expert knowledge or by statistical
testing, and inferring a (vine) copula for each separately, may reduce
this problem significantly. Additionally, vine inference on samples of
large size can become computationally demanding. Estimation techniques
based on parallel computing have been recently proposed to solve this
issue \citep{Wei2016_arXiv}. Additional work is foreseen to address these
challenges.

\section*{Acknowledgements}

The work has been funded by the ETH Foundation through the ETH Risk Center Seed Project SP-RC 07-15 ``Copulas for Big Data Analysis in Engineering Sciences'', and by the RiskLab from the Department of Mathematics of the ETH Zurich. The authors thank Moustapha Maliki for providing access to the Abaqus implementation of the dome model and to the UQLink module of $\uqlab$ needed to run it, and for helpful discussions on the interpretation of the dome response.

\bibliography{vines4uq_bibliography}

\appendix

\section{Some families of pair copulas and their properties} \label{sec:list_pair_copulas}

\tabref{pair_copula_cdfs} lists the $19$ parametric families of pair copulas implemented in the VineCopulaMatlab toolbox \citep{Kurz2014_CDVine} used here for vine inference. Each pair copula in the inferred vines was chosen among these families and their rotated versions defined by \eqrefp{rotated_copulas}, by selecting the family yielding the lowest AIC. The rotations of a pair-copula distribution $C$ are defined, here and in most references, by 
\begin{eqnarray}
C^{(90)}(u,v) &=& v - C(1-u, v), \nonumber \\  
C^{(180)}(u,v) &=& u+v-1 + C(1-u,1-v), \label{eq:rotated_copulas} \\
C^{(270)}(u,v) &=& u - C(u, 1-v). \nonumber 
\end{eqnarray}
(Note that $C^{(90)}$ and $C^{(270)}$ are obtained by flipping the copula density $c$ around the horizontal and vertical axis, respectively; some references provide the formulas for actual rotations: ${C^{(90)}(u,v)=v-C(v, 1-u)}$, ${C^{(270)}(u,v)=u-C(1-v,u)}$). Including the rotated copulas, $62$ families were considered in total for inference in our study. 

The analytical expressions for the Kendall's tau and for the coefficients $\lambda_l$, $\lambda_u$ of lower and upper tail dependence of the non-rotated families, when available, are reported in \tabref{pair_copula_properties}. We derived ourselves a few of these expressions, as indicated in the table, since we could not find them in the existing literature (see notes (a) and (c) in the table's caption). Note also that $\lambda_l$ and $\lambda_u$ switch when a copula density is rotated by $180^\circ$ and becomes its survival version. This allows copulas with lower tail dependence to be used to model upper tail dependence, and vice versa, by $180^\circ$ rotation. Copulas rotated by $90^\circ$ and $270^\circ$ model negative dependence.

\begin{table}[t!]
	\begin{center}
		\resizebox{\textwidth}{!}{
			\setlength{\extrarowheight}{1em}
			\begin{tabular}{llll}			
				\toprule
				ID & Name 
				& $\cdf$ 
				& Parameter range \\
				\cmidrule{1-4}
				1  & AMH
				& $\displaystyle{
					\frac{uv}{1-\theta(1-u)(1-v)}}$ 
				& $\theta \in [-1, 1]$ \\
				2  & AsymFGM 
				& $\displaystyle{uv \left( 1+\theta(1-u)^2 v (1-v) \right)}$ 
				& $\theta \in [0, 1]$ \\
				3  & BB1
				& $\displaystyle{
					\left( 1+ \left( (u^{-\theta_2}-1)^{\theta_1} + (v^{-\theta_2}-1)^{\theta_1} \right)^{1/\theta_1} \right)^{-1/\theta_2}}$
				& $\theta_1 \geq 1$, $\theta_2 > 0$  \\
				4  & BB6 
				& $\displaystyle{
					1 - \left(1-\exp\left\lbrace -\left[ (-\log(1-(1-u)^{\theta_2}))^{\theta_1} + (-\log(1-(1-v)^{\theta_2}))^{\theta_1} \right]^{1/{\theta_1}} \right\rbrace\right)^{1/\theta_2}}$ 
				& $\theta_1 \geq 1$, $\theta_2 \geq 1$  \\
				5  & BB7 
				& $\displaystyle{
					\varphi(\varphi^{-1}(u)+\varphi^{-1}(v))}$, where 
				$\varphi(w)=\varphi(w; \theta_1,\theta_2)=1-\left(1-(1+w)^{-1/\theta_1}\right)^{1/\theta_2}$ 
				& $\theta_1 \geq 1$, $\theta_2 >0$ \\
				6  & BB8 
				& $\displaystyle{\frac{1}{\theta_1}\left( 1 - \left( 1 - \frac{(1-(1-\theta_1 u)^{\theta_2}) 
						(1-(1-\theta_1 v)^{\theta_2})}{1-(1-\theta_1)^{\theta_2}}\right)^{1/\theta_2} \right)}$ 
				& $\theta_1 \geq 1$, $\theta_2 \in (0, 1]$ \\  
				7  & Clayton 
				& $\displaystyle{ 
					(u^{-\theta} + v^{-\theta} -1)^{-1/\theta}}$ 
				& $\theta > 0$ \\
				8  & FGM 
				& $\displaystyle{uv(1+\theta(1-u)(1-v))}$  
				& $\theta \in (-1, 1)$ \\                     
				9  & Frank 
				& $\displaystyle{-\frac{1}{\theta} \log \left( \frac{1-e^{-\theta} - (1-e^{-\theta u})(1-e^{-\theta v})}{1-e^{-\theta}} \right)}	$ 
				& $\theta \in \mathbb{R} \backslash \lbrace 0 \rbrace$  \\
				10 & Gaussian 
				& $\displaystyle{\Phi_{2; \theta} \left(\Phi^{-1}(u), \Phi^{-1}(v)\right)}$ $^{(a)}$(see \eqrefp{gaussian-copula}, with $d=2$) 
				& $\theta \in (-1, 1)$  \\      
				11 & Gumbel
				& $\exp\left( - ((-\log u)^\theta + (-\log v)^\theta)^{1/\theta} \right) $ 
				& $\theta \in [1, +\inf)$ \\
				12 & Iterated FGM 
				& $uv(1+\theta_1(1-u)(1-v) + \theta_2 uv(1-u)(1-v))$ 
				& $\theta_1, \theta_2 \in (-1, 1)$ \\  
				13 & Joe/B5 
				& $1-\left( (1-u)^\theta + (1-v)^\theta +(1-u)^\theta (1-v)^\theta  \right)^{1/\theta}$ 
				& $\theta \geq 1$ \\
				14 & Partial Frank 
				& $\displaystyle{\frac{uv}{\theta(u+v-uv)} (\log(1+(e^{-\theta}-1)(1+uv-u-v)) + \theta)}$
				& $\theta > 0$  \\[.75em]
				15 & Plackett 
				& $\displaystyle{\frac{1+(\theta-1)(u+v)- \sqrt{(1+(\theta-1)(u+v))^2-4\theta(\theta-1)uv}}{2(\theta-1)}}$
				& $\theta \geq 0$ \\
				16 & Tawn-1 
				& $\displaystyle{(uv)^{A\left( \frac{\log v}{\log(uv)}; \theta_1,\theta_3 \right)}}$, where 
				$\displaystyle{A(w; \theta_1,\theta_3)=(1-\theta_3)w+\left[ w^{\theta_1} + (\theta_3(1-w))^{\theta_1} \right]^{1/\theta_1}}$
				& $\theta_1 \geq 1$, $\theta_3 \in [0,1]$ \\
				17 & Tawn-2 
				& $\displaystyle{(uv)^{A\left( \frac{\log v}{\log(uv)}; \theta_1,\theta_2 \right)}}$, where 
				$\displaystyle{A(w; \theta_1,\theta_2)=(1-\theta_2)(1-w)+\left[ (\theta_2 w)^{\theta_1} + ((1-w))^{\theta_1} \right]^{1/\theta_1}}$
				& $\theta_1 \geq 1$, $\theta_2 \in [0,1]$ \\
				18 & Tawn 
				& $\displaystyle{(uv)^{A\left( w; \theta_1,\theta_2,\theta_3 \right)}}$, where $\displaystyle{w=\frac{\log v}{\log(uv)}}$ and \\
				& & $\displaystyle{A(w; \theta_1,\theta_2,\theta_3)=(1-\theta_2)(1-w)+(1-\theta_3)w+\left[ (\theta_2 w)^{\theta_1} + (\theta_3(1-w))^{\theta_1} \right]^{1/\theta_1}}$
				& $\theta_1 \geq 1$, $\theta_2,\theta_3 \in [0,1]$ \\
				19 & t- 
				& $\displaystyle{t_{2;\nu,\theta}\left( t_{\nu}^{-1}(u), t_{\nu}^{-1}(v) \right)} $ $^{(b)}$
				& $\nu>1$, $\theta \in (-1,1)$ \\[.5em]
				\bottomrule
			\end{tabular}
		}
		\caption{\textbf{Distributions of bivariate copula
                    families used for inference of vine copulas.} The
                  copula IDs are reported as assigned in the
                  VineCopulaMatlab toolbox used here
                  \citep{Kurz2014_CDVine}. $(a)$ $\Phi$ is the univariate
                  standard normal distribution, and $\Phi_{2;\theta}$ is
                  the bivariate normal distribution with zero means,
                  unit variance and correlation parameter $\theta$.
                  $(b)$ $t_{\nu}$ is the univariate $t$ distribution
                  with $\nu$ degrees of freedom, and $t_{\nu,\theta}$ is
                  the bivariate $t$ distribution with $\nu$ degrees of
                  freedom and correlation parameter $\theta$.}
		\label{tab:pair_copula_cdfs}
	\end{center}
\end{table}

\begin{table}[t!]
	\begin{center}
		\setlength{\extrarowheight}{1em}
		\resizebox{\textwidth}{!}{
			\begin{tabular}{llllll}
				\toprule
				ID & Name 
				& $\tau_K$
				& $\lambda_l$ \hspace{48pt}
				& $\lambda_u$
				& Special cases \\
				\cmidrule{1-6}
				1  & AMH
				& $\displaystyle{1 - \frac{2\theta + 2(1-\theta)^2 \ln(1-\theta)}{3\theta^2}}$ 
				& $\displaystyle{0.5 \cdot \bm{1}_{\lbrace \theta=1 \rbrace}}$ & $0$  
				& --- \\
				2  & AsymFGM 
				& $\displaystyle{\frac{\theta}{18}}$ $^{(a)}$
				& $0$ & $0$ 
				& --- \\
				3  & BB1 
				& $\displaystyle{1-\frac{2}{\theta_1(\theta_2+2)}}$ 
				& $\displaystyle{2^{-1/(\theta_1\theta_2)}}$ & $\displaystyle{2-2^{1/\theta_1}}$  
				& Clayton ($\theta_1=1$), \\[-1em]
				&&&&& Gumbel ($\theta_2 \downarrow 0^+$) \\
				4  & BB6 
				& numerical 
				& $0$ & $\displaystyle{2-2^{1/(\theta_1\theta_2)}}$  
				& Joe ($\theta_1=1$), \\[-1em]
				&&&&& Gumbel ($\theta_2=1$) \\
				5  & BB7 
				& see \citep{Schepsmeier2010_thesis}
				& $\displaystyle{2^{-1/\theta_1}}$ & $\displaystyle{2^{-1/\theta_2}}$  
				& Joe ($\theta_1 \downarrow 0^+$), \\[-1em]
				&&&&& Clayton ($\theta_2=1$)\\ 
				6  & BB8 
				& numerical  
				& $0$ & $0$ for $\theta_1\neq1$ 
				& Joe ($\theta_1 \downarrow 0^+$), \\[-1em]
				&&&&& Frank ($\theta_2 =1$)\\ 
				7  & Clayton 
				& $\displaystyle{\frac{\theta}{\theta+2}}$
				& $\displaystyle{2^{-1/\theta}}$ & $0$ 
				& ---\\ 
				8  & FGM 
				& $\displaystyle{\frac{2\theta}{9}}$
				& $0$ & $0$    
				& ---\\ 
				9  & Frank 
				& $\displaystyle{1+\frac{4}{\theta} (\frac{1}{\theta}\int_0^{\theta} t(e^t-1)^{-1}dt  - 1)}$ 
				& $0$ & $0$ 
				& ---\\ 
				10 & Gaussian 
				& $\displaystyle{\frac{2}{\pi} \arcsin(\theta)}$ 
				& $0$ & $0$ 
				& --- \\ 
				11 & Gumbel
				& $\displaystyle{\frac{\theta-1}{\theta}}$
				& $0$ & $\displaystyle{2-2^{1/\theta}}$
				& --- \\ 
				12 & Iterated FGM 
				& $\displaystyle{\frac{2\theta_1}{9} + \frac{(25+\theta_1)\theta_2}{450}}^{ \hspace{6pt}(a)}$ 
				& $0$ & $0$ 
				& FGM ($\theta_2=0$)\\ 
				13 & Joe/B5 
				& $\displaystyle{1+\frac{2}{2-\theta}(\digamma(2)-\digamma(\frac{2}{\theta}+1))}$ $^{(b)}$ 
				& $0$ & $\displaystyle{2-2^{1/\theta}}$
				& --- \\ 
				14 & Partial Frank \citep{Spanhel2016_76}
				& numerical
				& $0$ & $0$ 
				& --- \\
				15 & Plackett 
				& numerical 
				& $0$ & $0$
				& ---\\
				16 & Tawn-1 
				& numerical 
				& 0 $^{(c)}$ & $1+\theta_3 - \left( 1+\theta_3^{\theta_1}\right)^{1/\theta_1}$ $^{(c)}$
				& Gumbel ($\theta_3=1$) \\
				17 & Tawn-2 
				& numerical 
				& 0 $^{(c)}$ & $1+\theta_2 - \left( 1+\theta_2^{\theta_1}\right)^{1/\theta_1}$ $^{(c)}$
				& Gumbel ($\theta_2=1$) \\		   
				18 & Tawn 
				& numerical
				& 0 $^{(c)}$ & $\theta_2+\theta_3 - \left( \theta_2^{\theta_1} + \theta_3^{\theta_1}\right)^{1/\theta_1}$ $^{(c)}$
				& Tawn-1 ($\theta_2=1$), \\[-1em]
				&&&&& Tawn-2 ($\theta_3=1$), \\[-1em]
				&&&&& Gumbel ($\theta_2=\theta_3=1$) \\
				19 & t- 
				& $\displaystyle{\frac{2}{\pi} \arcsin(\theta)}$
				& $\lambda_l=\lambda_u=$ $^{(d)}$  &
				& --- \\[-1em]
				&&&\multicolumn{3}{c}{$\hspace{-24pt}=2t_{\nu+1} \left( -\sqrt{(\nu+1)(1-\theta)/(1+\theta)} \right)$} \\[.5em]
				\bottomrule
			\end{tabular}
		}
		\caption{\textbf{Some properties of the considered pair copulas.} Kendall's tau, tail dependence coefficients, subfamilies of pair copulas that obtain for specific parameter values. $(a)$ We derived the analytical expression of $\tau_K$ for the asymmetric and iterated FGM copulas using the RHS of \eqrefp{Kendalls-tau-from-C}. ${(b)}$ $\digamma$ is the digamma function. $(c)$ We derived the analytical expression of the tail dependence coefficients by using \eqrefp{tail_dependence_U}, by noting that $A(w)=1+\frac{1}{2}\left( (\theta_2^{\theta_1}+\theta_3^{\theta_1} )^{1/\theta_1} - (\theta_2+\theta_3) \right)$ when $u=v$ and, for $\lambda_u$, by calculating the limit through first order Taylor expansion. $(d)$ $t_{\nu}$ is the univariate $t$ distribution with $\nu$ degrees of freedom.}
		\label{tab:pair_copula_properties}
	\end{center}
\end{table}

\end{document}